
%
%
%
%
%
%
%
%
\def\standardrisposta{s }\def\reducedrisposta{r }
\def\mplarisposta{mpla }\def\zerorisposta{z }
\def\doublerisposta{d }\def\cartarisposta{e }\def\amsrisposta{y }
\newcount\ingrandimento \newcount\sinnota \newcount\dimnota
\newcount\unoduecol \newdimen\collhsize \newdimen\tothsize
\newdimen\fullhsize \newcount\controllorisposta \sinnota=1
\newskip\infralinea  \global\controllorisposta=0
\immediate\write16 { ********  Welcome to PANDA macros (Plain TeX,
AP, 1991) ******** }
\immediate\write16 { You'll have to answer a few questions in
lowercase.}
\message{>  Do you want it in double-page (d), reduced (r)
or standard format (s) ? }\read-1 to\risposta
\message{>  Do you want it in USA A4 (u) or EUROPEAN A4
(e) paper size ? }\read-1 to\srisposta
\message{>  Do you have AMSFonts 2.0 (math) fonts (y/n) ? }
\read-1 to\arisposta
%
%
%
%
%
\ifx\risposta\standardrisposta \ingrandimento=1200
\message {>> This will come out UNREDUCED << }
\dimnota=2 \unoduecol=1 \global\controllorisposta=1 \fi
\ifx\risposta\reducedrisposta \ingrandimento=1095 \dimnota=1
\unoduecol=1  \global\controllorisposta=1
\message {>> This will come out REDUCED << } \fi
\ifx\risposta\doublerisposta \ingrandimento=1000 \dimnota=2
\unoduecol=2   \message {>> You must print this in
LANDSCAPE orientation << } \global\controllorisposta=1 \fi
\ifx\risposta\mplarisposta \ingrandimento=1000 \dimnota=1
\message {>> Mod. Phys. Lett. A format << }
\unoduecol=1 \global\controllorisposta=1 \fi
\ifx\risposta\zerorisposta \ingrandimento=1000 \dimnota=2
\message {>> Zero Magnification format << }
\unoduecol=1 \global\controllorisposta=1 \fi
\ifnum\controllorisposta=0  \ingrandimento=1200
\message {>>> ERROR IN INPUT, I ASSUME STANDARD
UNREDUCED FORMAT <<< }  \dimnota=2 \unoduecol=1 \fi
\magnification=\ingrandimento
%
%
%
%
\newdimen\eucolumnsize \newdimen\eudoublehsize \newdimen\eudoublevsize
\newdimen\uscolumnsize \newdimen\usdoublehsize \newdimen\usdoublevsize
\newdimen\eusinglehsize \newdimen\eusinglevsize \newdimen\ussinglehsize
\newskip\standardbaselineskip \newdimen\ussinglevsize
\newskip\reducedbaselineskip \newskip\doublebaselineskip
\eucolumnsize=12.0truecm    
\eudoublehsize=25.5truecm   
\eudoublevsize=6.5truein    
\uscolumnsize=4.4truein     
\usdoublehsize=9.4truein    
\usdoublevsize=6.8truein    
\eusinglehsize=6.5truein    
\eusinglevsize=24truecm     
\ussinglehsize=6.5truein    
\ussinglevsize=8.9truein    
\standardbaselineskip=16pt plus.2pt  
\reducedbaselineskip=14pt plus.2pt   
\doublebaselineskip=12pt plus.2pt    
%
%
\def\Portoffset{}
\def\Landoffset{}
\ifx\risposta\mplarisposta \def\Portoffset{\hoffset=1.8truecm} \fi
%
%
\def\Landspec{}
\tolerance=10000
\parskip=0pt plus2pt  \leftskip=0pt \rightskip=0pt
%
%
\ifx\risposta\standardrisposta \infralinea=\standardbaselineskip \fi
\ifx\risposta\reducedrisposta  \infralinea=\reducedbaselineskip \fi
\ifx\risposta\doublerisposta   \infralinea=\doublebaselineskip \fi
\ifx\risposta\mplarisposta     \infralinea=13pt \fi
\ifx\risposta\zerorisposta     \infralinea=12pt plus.2pt\fi
\ifnum\controllorisposta=0    \infralinea=\standardbaselineskip \fi
\ifx\risposta\doublerisposta   \Landoffset \else \Portoffset \fi
\ifx\risposta\doublerisposta \ifx\srisposta\cartarisposta
\tothsize=\eudoublehsize \collhsize=\eucolumnsize
\vsize=\eudoublevsize  \else  \tothsize=\usdoublehsize
\collhsize=\uscolumnsize \vsize=\usdoublevsize \fi \else
\ifx\srisposta\cartarisposta \tothsize=\eusinglehsize
\vsize=\eusinglevsize \else  \tothsize=\ussinglehsize
\vsize=\ussinglevsize \fi \collhsize=4.4truein \fi
\ifx\risposta\mplarisposta \tothsize=5.0truein
\vsize=7.8truein \collhsize=4.4truein \fi
%
%
%
%
\newcount\contaeuler \newcount\contacyrill \newcount\contaams
\font\ninerm=cmr9  \font\eightrm=cmr8  \font\sixrm=cmr6
\font\ninei=cmmi9  \font\eighti=cmmi8  \font\sixi=cmmi6
\font\ninesy=cmsy9  \font\eightsy=cmsy8  \font\sixsy=cmsy6
\font\ninebf=cmbx9  \font\eightbf=cmbx8  \font\sixbf=cmbx6
\font\ninett=cmtt9  \font\eighttt=cmtt8  \font\nineit=cmti9
\font\eightit=cmti8 \font\ninesl=cmsl9  \font\eightsl=cmsl8
\skewchar\ninei='177 \skewchar\eighti='177 \skewchar\sixi='177
\skewchar\ninesy='60 \skewchar\eightsy='60 \skewchar\sixsy='60
\hyphenchar\ninett=-1 \hyphenchar\eighttt=-1 \hyphenchar\tentt=-1
%
\font\tencmmib=cmmib10  \newfam\cmmibfam  \skewchar\tencmmib='177
\font\tencmbsy=cmbsy10  \newfam\cmbsyfam  \skewchar\tencmbsy='60
\def\scaps{\cmcsc}                 
\font\tencmcsc=cmcsc10  \newfam\cmcscfam
\ifnum\ingrandimento=1095

\font\capsone=cmcsc10 at 10.95pt 

\else

\font\capsone=cmcsc10 at 12pt 
\fi

\def\ttaarr{\bf}		
\def\ppaarr{\sl}		

%
%
%
\newfam\eufmfam \newfam\msamfam \newfam\msbmfam \newfam\eufbfam
\def\Loadeulerfonts{\global\contaeuler=1 \ifx\arisposta\amsrisposta
\font\teneufm=eufm10              
\font\eighteufm=eufm8 \font\nineeufm=eufm9 \font\sixeufm=eufm6
\font\seveneufm=eufm7  \font\fiveeufm=eufm5
\font\teneufb=eufb10              
\font\eighteufb=eufb8 \font\nineeufb=eufb9 \font\sixeufb=eufb6
\font\seveneufb=eufb7  \font\fiveeufb=eufb5
\font\teneurm=eurm10              
\font\eighteurm=eurm8 \font\nineeurm=eurm9
\font\teneurb=eurb10              
\font\eighteurb=eurb8 \font\nineeurb=eurb9
\font\teneusm=eusm10              
\font\eighteusm=eusm8 \font\nineeusm=eusm9
\font\teneusb=eusb10              
\font\eighteusb=eusb8 \font\nineeusb=eusb9
\else \def\eufm{\tt} \def\eufb{\tt} \def\eurm{\tt} \def\eurb{\tt}
\def\eusm{\tt} \def\eusb{\tt}    \fi}

\def\loadamsmath{\global\contaams=1 \ifx\arisposta\amsrisposta
\font\tenmsam=msam10 \font\ninemsam=msam9 \font\eightmsam=msam8
\font\sevenmsam=msam7 \font\sixmsam=msam6 \font\fivemsam=msam5
\font\tenmsbm=msbm10 \font\ninemsbm=msbm9 \font\eightmsbm=msbm8
\font\sevenmsbm=msbm7 \font\sixmsbm=msbm6 \font\fivemsbm=msbm5
\else \def\msbm{\bf} \fi \def\Bbb{\msbm} \def\symbl{\msam} \tenpoint}
\def\loadcyrill{\global\contacyrill=1 \ifx\arisposta\amsrisposta
\font\tenwncyr=wncyr10 \font\ninewncyr=wncyr9 \font\eightwncyr=wncyr8
\font\tenwncyb=wncyr10 \font\ninewncyb=wncyr9 \font\eightwncyb=wncyr8
\font\tenwncyi=wncyr10 \font\ninewncyi=wncyr9 \font\eightwncyi=wncyr8
\else \def\cyrill{\sl} \def\cyrilb{\sl} \def\cyrili{\sl} \fi\tenpoint}
\ifx\arisposta\amsrisposta
\font\sevenex=cmex7               
\font\eightex=cmex8  \font\nineex=cmex9
\font\ninecmmib=cmmib9   \font\eightcmmib=cmmib8
\font\sevencmmib=cmmib7 \font\sixcmmib=cmmib6
\font\fivecmmib=cmmib5   \skewchar\ninecmmib='177
\skewchar\eightcmmib='177  \skewchar\sevencmmib='177
\skewchar\sixcmmib='177   \skewchar\fivecmmib='177
\font\ninecmbsy=cmbsy9    \font\eightcmbsy=cmbsy8
\font\sevencmbsy=cmbsy7  \font\sixcmbsy=cmbsy6
\font\fivecmbsy=cmbsy5   \skewchar\ninecmbsy='60
\skewchar\eightcmbsy='60  \skewchar\sevencmbsy='60
\skewchar\sixcmbsy='60    \skewchar\fivecmbsy='60
\font\ninecmcsc=cmcsc9    \font\eightcmcsc=cmcsc8     \else
\def\cmmib{\fam\cmmibfam\tencmmib}\textfont\cmmibfam=\tencmmib
\scriptfont\cmmibfam=\tencmmib \scriptscriptfont\cmmibfam=\tencmmib
\def\cmbsy{\fam\cmbsyfam\tencmbsy} \textfont\cmbsyfam=\tencmbsy
\scriptfont\cmbsyfam=\tencmbsy \scriptscriptfont\cmbsyfam=\tencmbsy
\scriptfont\cmcscfam=\tencmcsc \scriptscriptfont\cmcscfam=\tencmcsc
\def\cmcsc{\fam\cmcscfam\tencmcsc} \textfont\cmcscfam=\tencmcsc \fi
\catcode`@=11
\newskip\ttglue
\gdef\tenpoint{\def\rm{\fam0\tenrm}
  \textfont0=\tenrm \scriptfont0=\sevenrm \scriptscriptfont0=\fiverm
  \textfont1=\teni \scriptfont1=\seveni \scriptscriptfont1=\fivei
  \textfont2=\tensy \scriptfont2=\sevensy \scriptscriptfont2=\fivesy
  \textfont3=\tenex \scriptfont3=\tenex \scriptscriptfont3=\tenex
  \def\mcal{\fam2 \tensy}  \def\mmit{\fam1 \teni}
  \textfont\itfam=\tenit \def\it{\fam\itfam\tenit}
  \textfont\slfam=\tensl \def\sl{\fam\slfam\tensl}
  \textfont\ttfam=\tentt \scriptfont\ttfam=\eighttt
  \scriptscriptfont\ttfam=\eighttt  \def\tt{\fam\ttfam\tentt}
  \textfont\bffam=\tenbf \scriptfont\bffam=\sevenbf
  \scriptscriptfont\bffam=\fivebf \def\bf{\fam\bffam\tenbf}
     \ifx\arisposta\amsrisposta    \ifnum\contaeuler=1
  \textfont\eufmfam=\teneufm \scriptfont\eufmfam=\seveneufm
  \scriptscriptfont\eufmfam=\fiveeufm \def\eufm{\fam\eufmfam\teneufm}
  \textfont\eufbfam=\teneufb \scriptfont\eufbfam=\seveneufb
  \scriptscriptfont\eufbfam=\fiveeufb \def\eufb{\fam\eufbfam\teneufb}
  \def\eurm{\teneurm} \def\eurb{\teneurb} \def\eusm{\teneusm}
  \def\eusb{\teneusb}    \fi    \ifnum\contaams=1
  \textfont\msamfam=\tenmsam \scriptfont\msamfam=\sevenmsam
  \scriptscriptfont\msamfam=\fivemsam \def\msam{\fam\msamfam\tenmsam}
  \textfont\msbmfam=\tenmsbm \scriptfont\msbmfam=\sevenmsbm
  \scriptscriptfont\msbmfam=\fivemsbm \def\msbm{\fam\msbmfam\tenmsbm}
     \fi      \ifnum\contacyrill=1     \def\cyrill{\tenwncyr}
  \def\cyrilb{\tenwncyb}  \def\cyrili{\tenwncyi}         \fi
  \textfont3=\tenex \scriptfont3=\sevenex \scriptscriptfont3=\sevenex
  \def\cmmib{\fam\cmmibfam\tencmmib} \scriptfont\cmmibfam=\sevencmmib
  \textfont\cmmibfam=\tencmmib  \scriptscriptfont\cmmibfam=\fivecmmib
  \def\cmbsy{\fam\cmbsyfam\tencmbsy} \scriptfont\cmbsyfam=\sevencmbsy
  \textfont\cmbsyfam=\tencmbsy  \scriptscriptfont\cmbsyfam=\fivecmbsy
  \def\cmcsc{\fam\cmcscfam\tencmcsc} \scriptfont\cmcscfam=\eightcmcsc
  \textfont\cmcscfam=\tencmcsc \scriptscriptfont\cmcscfam=\eightcmcsc
     \fi            \tt \ttglue=.5em plus.25em minus.15em
  \normalbaselineskip=12pt
  \setbox\strutbox=\hbox{\vrule height8.5pt depth3.5pt width0pt}
  \let\sc=\eightrm \let\big=\tenbig   \normalbaselines
  \baselineskip=\infralinea  \rm}
\gdef\ninepoint{\def\rm{\fam0\ninerm}
  \textfont0=\ninerm \scriptfont0=\sixrm \scriptscriptfont0=\fiverm
  \textfont1=\ninei \scriptfont1=\sixi \scriptscriptfont1=\fivei
  \textfont2=\ninesy \scriptfont2=\sixsy \scriptscriptfont2=\fivesy
  \textfont3=\tenex \scriptfont3=\tenex \scriptscriptfont3=\tenex
  \def\mcal{\fam2 \ninesy}  \def\mmit{\fam1 \ninei}
  \textfont\itfam=\nineit \def\it{\fam\itfam\nineit}
  \textfont\slfam=\ninesl \def\sl{\fam\slfam\ninesl}
  \textfont\ttfam=\ninett \scriptfont\ttfam=\eighttt
  \scriptscriptfont\ttfam=\eighttt \def\tt{\fam\ttfam\ninett}
  \textfont\bffam=\ninebf \scriptfont\bffam=\sixbf
  \scriptscriptfont\bffam=\fivebf \def\bf{\fam\bffam\ninebf}
     \ifx\arisposta\amsrisposta  \ifnum\contaeuler=1
  \textfont\eufmfam=\nineeufm \scriptfont\eufmfam=\sixeufm
  \scriptscriptfont\eufmfam=\fiveeufm \def\eufm{\fam\eufmfam\nineeufm}
  \textfont\eufbfam=\nineeufb \scriptfont\eufbfam=\sixeufb
  \scriptscriptfont\eufbfam=\fiveeufb \def\eufb{\fam\eufbfam\nineeufb}
  \def\eurm{\nineeurm} \def\eurb{\nineeurb} \def\eusm{\nineeusm}
  \def\eusb{\nineeusb}     \fi   \ifnum\contaams=1
  \textfont\msamfam=\ninemsam \scriptfont\msamfam=\sixmsam
  \scriptscriptfont\msamfam=\fivemsam \def\msam{\fam\msamfam\ninemsam}
  \textfont\msbmfam=\ninemsbm \scriptfont\msbmfam=\sixmsbm
  \scriptscriptfont\msbmfam=\fivemsbm \def\msbm{\fam\msbmfam\ninemsbm}
     \fi       \ifnum\contacyrill=1     \def\cyrill{\ninewncyr}
  \def\cyrilb{\ninewncyb}  \def\cyrili{\ninewncyi}         \fi
  \textfont3=\nineex \scriptfont3=\sevenex \scriptscriptfont3=\sevenex
  \def\cmmib{\fam\cmmibfam\ninecmmib}  \textfont\cmmibfam=\ninecmmib
  \scriptfont\cmmibfam=\sixcmmib \scriptscriptfont\cmmibfam=\fivecmmib
  \def\cmbsy{\fam\cmbsyfam\ninecmbsy}  \textfont\cmbsyfam=\ninecmbsy
  \scriptfont\cmbsyfam=\sixcmbsy \scriptscriptfont\cmbsyfam=\fivecmbsy
  \def\cmcsc{\fam\cmcscfam\ninecmcsc} \scriptfont\cmcscfam=\eightcmcsc
  \textfont\cmcscfam=\ninecmcsc \scriptscriptfont\cmcscfam=\eightcmcsc
     \fi            \tt \ttglue=.5em plus.25em minus.15em
  \normalbaselineskip=11pt
  \setbox\strutbox=\hbox{\vrule height8pt depth3pt width0pt}
  \let\sc=\sevenrm \let\big=\ninebig \normalbaselines\rm}
\gdef\eightpoint{\def\rm{\fam0\eightrm}
  \textfont0=\eightrm \scriptfont0=\sixrm \scriptscriptfont0=\fiverm
  \textfont1=\eighti \scriptfont1=\sixi \scriptscriptfont1=\fivei
  \textfont2=\eightsy \scriptfont2=\sixsy \scriptscriptfont2=\fivesy
  \textfont3=\tenex \scriptfont3=\tenex \scriptscriptfont3=\tenex
  \def\mcal{\fam2 \eightsy}  \def\mmit{\fam1 \eighti}
  \textfont\itfam=\eightit \def\it{\fam\itfam\eightit}
  \textfont\slfam=\eightsl \def\sl{\fam\slfam\eightsl}
  \textfont\ttfam=\eighttt \scriptfont\ttfam=\eighttt
  \scriptscriptfont\ttfam=\eighttt \def\tt{\fam\ttfam\eighttt}
  \textfont\bffam=\eightbf \scriptfont\bffam=\sixbf
  \scriptscriptfont\bffam=\fivebf \def\bf{\fam\bffam\eightbf}
     \ifx\arisposta\amsrisposta   \ifnum\contaeuler=1
  \textfont\eufmfam=\eighteufm \scriptfont\eufmfam=\sixeufm
  \scriptscriptfont\eufmfam=\fiveeufm \def\eufm{\fam\eufmfam\eighteufm}
  \textfont\eufbfam=\eighteufb \scriptfont\eufbfam=\sixeufb
  \scriptscriptfont\eufbfam=\fiveeufb \def\eufb{\fam\eufbfam\eighteufb}
  \def\eurm{\eighteurm} \def\eurb{\eighteurb} \def\eusm{\eighteusm}
  \def\eusb{\eighteusb}       \fi    \ifnum\contaams=1
  \textfont\msamfam=\eightmsam \scriptfont\msamfam=\sixmsam
  \scriptscriptfont\msamfam=\fivemsam \def\msam{\fam\msamfam\eightmsam}
  \textfont\msbmfam=\eightmsbm \scriptfont\msbmfam=\sixmsbm
  \scriptscriptfont\msbmfam=\fivemsbm \def\msbm{\fam\msbmfam\eightmsbm}
     \fi       \ifnum\contacyrill=1     \def\cyrill{\eightwncyr}
  \def\cyrilb{\eightwncyb}  \def\cyrili{\eightwncyi}         \fi
  \textfont3=\eightex \scriptfont3=\sevenex \scriptscriptfont3=\sevenex
  \def\cmmib{\fam\cmmibfam\eightcmmib}  \textfont\cmmibfam=\eightcmmib
  \scriptfont\cmmibfam=\sixcmmib \scriptscriptfont\cmmibfam=\fivecmmib
  \def\cmbsy{\fam\cmbsyfam\eightcmbsy}  \textfont\cmbsyfam=\eightcmbsy
  \scriptfont\cmbsyfam=\sixcmbsy \scriptscriptfont\cmbsyfam=\fivecmbsy
  \def\cmcsc{\fam\cmcscfam\eightcmcsc} \scriptfont\cmcscfam=\eightcmcsc
  \textfont\cmcscfam=\eightcmcsc \scriptscriptfont\cmcscfam=\eightcmcsc
     \fi             \tt \ttglue=.5em plus.25em minus.15em
  \normalbaselineskip=9pt
  \setbox\strutbox=\hbox{\vrule height7pt depth2pt width0pt}
  \let\sc=\sixrm \let\big=\eightbig \normalbaselines\rm }
\gdef\tenbig#1{{\hbox{$\left#1\vbox to8.5pt{}\right.\n@space$}}}
\gdef\ninebig#1{{\hbox{$\textfont0=\tenrm\textfont2=\tensy
   \left#1\vbox to7.25pt{}\right.\n@space$}}}
\gdef\eightbig#1{{\hbox{$\textfont0=\ninerm\textfont2=\ninesy
   \left#1\vbox to6.5pt{}\right.\n@space$}}}
\def\alternativefont#1#2{\ifx\arisposta\amsrisposta \relax \else
\xdef#1{#2} \fi}
\global\contaeuler=0 \global\contacyrill=0 \global\contaams=0
%
%
%
%
\newbox\fotlinebb \newbox\hedlinebb \newbox\leftcolumn
\gdef\makeheadline{\vbox to 0pt{\vskip-22.5pt
     \fullline{\vbox to8.5pt{}\the\headline}\vss}\nointerlineskip}
\gdef\makehedlinebb{\vbox to 0pt{\vskip-22.5pt
     \fullline{\vbox to8.5pt{}\copy\hedlinebb\hfil
     \line{\hfill\the\headline\hfill}}\vss} \nointerlineskip}
\gdef\makefootline{\baselineskip=24pt \fullline{\the\footline}}
\gdef\makefotlinebb{\baselineskip=24pt
    \fullline{\copy\fotlinebb\hfil\line{\hfill\the\footline\hfill}}}
\gdef\doubleformat{\shipout\vbox{\Landspec\makehedlinebb
     \fullline{\box\leftcolumn\hfil\columnbox}\makefotlinebb}
     \advancepageno}
\gdef\columnbox{\leftline{\pagebody}}
\gdef\line#1{\hbox to\hsize{\hskip\leftskip#1\hskip\rightskip}}
\gdef\fullline#1{\hbox to\fullhsize{\hskip\leftskip{#1}%
\hskip\rightskip}}
\gdef\footnote#1{\let\@sf=\empty
         \ifhmode\edef\#sf{\spacefactor=\the\spacefactor}\/\fi
         #1\@sf\vfootnote{#1}}
\gdef\vfootnote#1{\insert\footins\bgroup
         \ifnum\dimnota=1  \eightpoint\fi
         \ifnum\dimnota=2  \ninepoint\fi
         \ifnum\dimnota=0  \tenpoint\fi
         \interlinepenalty=\interfootnotelinepenalty
         \splittopskip=\ht\strutbox
         \splitmaxdepth=\dp\strutbox \floatingpenalty=20000
         \leftskip=\oldssposta \rightskip=\olddsposta
         \spaceskip=0pt \xspaceskip=0pt
         \ifnum\sinnota=0   \textindent{#1}\fi
         \ifnum\sinnota=1   \item{#1}\fi
         \footstrut\futurelet\next\fo@t}
\gdef\fo@t{\ifcat\bgroup\noexpand\next \let\next\f@@t
             \else\let\next\f@t\fi \next}
\gdef\f@@t{\bgroup\aftergroup\@foot\let\next}
\gdef\f@t#1{#1\@foot} \gdef\@foot{\strut\egroup}
\gdef\footstrut{\vbox to\splittopskip{}}
\skip\footins=\bigskipamount
\count\footins=1000  \dimen\footins=8in
\catcode`@=12
\tenpoint
\ifnum\unoduecol=1 \hsize=\tothsize   \fullhsize=\tothsize \fi
\ifnum\unoduecol=2 \hsize=\collhsize  \fullhsize=\tothsize \fi
\global\let\lrcol=L      \ifnum\unoduecol=1
\output{\plainoutput{\ifnum\tipbnota=2 \clearnmbnota\fi}} \fi
\ifnum\unoduecol=2 \output{\if L\lrcol
     \global\setbox\leftcolumn=\columnbox
     \global\setbox\fotlinebb=\line{\hfill\the\footline\hfill}
     \global\setbox\hedlinebb=\line{\hfill\the\headline\hfill}
     \advancepageno  \global\let\lrcol=R
     \else  \doubleformat \global\let\lrcol=L \fi
     \ifnum\outputpenalty>-20000 \else\dosupereject\fi
     \ifnum\tipbnota=2\clearnmbnota\fi }\fi
\def\ifdoublepage{\ifnum\unoduecol=2 }
\gdef\yespagenumbers{\footline={\hss\tenrm\folio\hss}}
\gdef\ciao{ \ifnum\fdefcontre=1 \endfdef\fi
     \par\vfill\supereject \ifnum\unoduecol=2
     \if R\lrcol  \headline={}\nopagenumbers\null\vfill\eject
     \fi\fi \end}

\newskip\olddsposta \newskip\oldssposta
\global\oldssposta=\leftskip \global\olddsposta=\rightskip

\def\filldots{\leaders\hbox to 1em{\hss.\hss}\hfill}
\def\inquadrb#1 {\vbox {\hrule  \hbox{\vrule \vbox {\vskip .2cm
    \hbox {\ #1\ } \vskip .2cm } \vrule  }  \hrule} }
 \def\newline{\hfil\break}
\def\jump{\vskip\baselineskip} \newskip\iinnffrr
\def\sjump{\iinnffrr=\baselineskip
          \divide\iinnffrr by 2 \vskip\iinnffrr}
\def\bjump{\vskip\baselineskip \vskip\baselineskip}
\newcount\nmbnota  \def\clearnmbnota{\global\nmbnota=0}
\newcount\tipbnota \def\letterfootnote{\global\tipbnota=1}

\def\note#1{\global\advance\nmbnota by 1 \ifnum\tipbnota=1
    \footnote{$^{\rm\nttlett}$}{#1} \else {\ifnum\tipbnota=2
    \footnote{$^{\nttsymb}$}{#1}
    \else\footnote{$^{\the\nmbnota}$}{#1}\fi}\fi}
\def\nttlett{\ifcase\nmbnota \or a\or b\or c\or d\or e\or f\or
g\or h\or i\or j\or k\or l\or m\or n\or o\or p\or q\or r\or
s\or t\or u\or v\or w\or y\or x\or z\fi}
\def\nttsymb{\ifcase\nmbnota \or\dag\or\sharp\or\ddag\or\star\or
\natural\or\flat\or\clubsuit\or\diamondsuit\or\heartsuit
\or\spadesuit\fi}   \clearnmbnota
\def\numberfootnote{\global\tipbnota=0} \numberfootnote
\def\setnote#1{\expandafter\xdef\csname#1\endcsname{
\ifnum\tipbnota=1 {\rm\nttlett} \else {\ifnum\tipbnota=2
{\nttsymb} \else \the\nmbnota\fi}\fi} }
\newcount\nbmfig  \def\clearnbmfig{\global\nbmfig=0}
\gdef\figure{\global\advance\nbmfig by 1
      {\rm fig. \the\nbmfig}}   \clearnbmfig
\def\setfig#1{\expandafter\xdef\csname#1\endcsname{fig. \the\nbmfig}}
 \def\endformula{\eqno\numero $$}
 \def\efr{\endformula}
\newcount\frmcount \def\clearfrmcount{\global\frmcount=0}
\def\numero{\global\advance\frmcount by 1   \ifnum\indappcount=0
  {\ifnum\cpcount <1 {\hbox{\rm (\the\frmcount )}}  \else
  {\hbox{\rm (\the\cpcount .\the\frmcount )}} \fi}  \else
  {\hbox{\rm (\applett .\the\frmcount )}} \fi}
\def\nameformula#1{\global\advance\frmcount by 1%
\ifnum\draftnum=0  {\ifnum\indappcount=0%
{\ifnum\cpcount<1\xdef\spzzttrra{(\the\frmcount )}%
\else\xdef\spzzttrra{(\the\cpcount .\the\frmcount )}\fi}%
\else\xdef\spzzttrra{(\applett .\the\frmcount )}\fi}%
\else\xdef\spzzttrra{(#1)}\fi%
\expandafter\xdef\csname#1\endcsname{\spzzttrra}
\eqno \hbox{\rm\spzzttrra} $$}
\def\nfr{\nameformula}    
\def\nameali#1{\global\advance\frmcount by 1%
\ifnum\draftnum=0  {\ifnum\indappcount=0%
{\ifnum\cpcount<1\xdef\spzzttrra{(\the\frmcount )}%
\else\xdef\spzzttrra{(\the\cpcount .\the\frmcount )}\fi}%
\else\xdef\spzzttrra{(\applett .\the\frmcount )}\fi}%
\else\xdef\spzzttrra{(#1)}\fi%
\expandafter\xdef\csname#1\endcsname{\spzzttrra}
  \hbox{\rm\spzzttrra} }      \clearfrmcount
\newcount\cpcount \def\clearcpcount{\global\cpcount=0}
\newcount\subcpcount \def\clearsubcpcount{\global\subcpcount=0}
\newcount\appcount \def\clearappcount{\global\appcount=0}
\newcount\indappcount \def\clearindappcount{\indappcount=0}
\newcount\sottoparcount 

\def\applett{\ifcase\appcount  \or {A}\or {B}\or {C}\or
{D}\or {E}\or {F}\or {G}\or {H}\or {I}\or {J}\or {K}\or {L}\or
{M}\or {N}\or {O}\or {P}\or {Q}\or {R}\or {S}\or {T}\or {U}\or
{V}\or {W}\or {X}\or {Y}\or {Z}\fi    \ifnum\appcount<0
\immediate\write16 {Panda ERROR - Appendix: counter "appcount"
out of range}\fi  \ifnum\appcount>26  \immediate\write16 {Panda
ERROR - Appendix: counter "appcount" out of range}\fi}
\clearappcount  \clearindappcount \newcount\connttrre
\def\clearconnttrre{\global\connttrre=0} \newcount\countref
\def\clearcountref{\global\countref=0} \clearcountref
\def\chapter#1{\global\advance\cpcount by 1 \clearfrmcount
                 \goodbreak\null\vbox{\jump\nobreak
                 \clearsubcpcount\clearindappcount
                 \itemitem{\ttaarr\the\cpcount .\qquad}{\ttaarr #1}
                 \par\nobreak\jump\sjump}\nobreak}
\def\section#1{\global\advance\subcpcount by 1 \goodbreak\null
               \vbox{\sjump\nobreak\ifnum\indappcount=0
                 {\ifnum\cpcount=0 {\itemitem{\ppaarr
               .\the\subcpcount\quad\enskip\ }{\ppaarr #1}\par} \else
                 {\itemitem{\ppaarr\the\cpcount .\the\subcpcount\quad
                  \enskip\ }{\ppaarr #1} \par}  \fi}
                \else{\itemitem{\ppaarr\applett .\the\subcpcount\quad
                 \enskip\ }{\ppaarr #1}\par}\fi\nobreak\jump}\nobreak}
\clearsubcpcount
\def\appendix#1{\global\advance\appcount by 1 \clearfrmcount
                  \goodbreak\null\vbox{\jump\nobreak
                  \global\advance\indappcount by 1 \clearsubcpcount
          \itemitem{ }{\hskip-40pt\ttaarr Appendix\ #1}
             \nobreak\jump\sjump}\nobreak}
\clearappcount \clearindappcount
\def\references{\goodbreak\null\vbox{\jump\nobreak
   \itemitem{}{\ttaarr References} \nobreak\jump\sjump}\nobreak}

\clearcpcount\clearcountref

\def\setchap#1{\ifnum\indappcount=0{\ifnum\subcpcount=0%
\xdef\spzzttrra{\the\cpcount}%
\else\xdef\spzzttrra{\the\cpcount .\the\subcpcount}\fi}
\else{\ifnum\subcpcount=0 \xdef\spzzttrra{\applett}%
\else\xdef\spzzttrra{\applett .\the\subcpcount}\fi}\fi
\expandafter\xdef\csname#1\endcsname{\spzzttrra}}
\newcount\draftnum \newcount\ppora   \newcount\ppminuti
\global\ppora=\time   \global\ppminuti=\time
\global\divide\ppora by 60  \draftnum=\ppora
\multiply\draftnum by 60    \global\advance\ppminuti by -\draftnum
\def\droggi{\number\day /\number\month /\number\year\ \the\ppora
:\the\ppminuti}     \global\draftnum=0
\def\draftcomment#1{\ifnum\draftnum=0 \relax \else
{\ {\bf ***}\ #1\ {\bf ***}\ }\fi} 
%
%
\catcode`@=11
\gdef\Ref#1{\expandafter\ifx\csname @rrxx@#1\endcsname\relax%
{\global\advance\countref by 1    \ifnum\countref>200
\immediate\write16 {Panda ERROR - Ref: maximum number of references
exceeded}  \expandafter\xdef\csname @rrxx@#1\endcsname{0}\else
\expandafter\xdef\csname @rrxx@#1\endcsname{\the\countref}\fi}\fi
\ifnum\draftnum=0 \csname @rrxx@#1\endcsname \else#1\fi}
\gdef\beginref{\ifnum\draftnum=0  \gdef\Rref{\fairef}
\gdef\endref{\scriviref} \else\relax\fi
\ifx\risposta\mplarisposta \ninepoint \fi
\parskip 2pt plus.2pt \baselineskip=12pt}
\def\Reflab#1{[#1]} \gdef\Rref#1#2{\item{\Reflab{#1}}{#2}}
\gdef\endref{\relax}  \newcount\conttemp
\gdef\fairef#1#2{\expandafter\ifx\csname @rrxx@#1\endcsname\relax
{\global\conttemp=0 \immediate\write16 {Panda ERROR - Ref: reference
[#1] undefined}} \else
{\global\conttemp=\csname @rrxx@#1\endcsname } \fi
\global\advance\conttemp by 50  \global\setbox\conttemp=\hbox{#2} }
\gdef\scriviref{\clearconnttrre\conttemp=50
\loop\ifnum\connttrre<\countref \advance\conttemp by 1
\advance\connttrre by 1
\item{\Reflab{\the\connttrre}}{\unhcopy\conttemp} \repeat}
\clearcountref \clearconnttrre
\catcode`@=12
\ifx\risposta\mplarisposta \def\Reflab#1{#1.} \letterfootnote \fi

\def\slashchar#1{\setbox0=\hbox{$#1$} \dimen0=\wd0
     \setbox1=\hbox{/} \dimen1=\wd1 \ifdim\dimen0>\dimen1
      \rlap{\hbox to \dimen0{\hfil/\hfil}} #1 \else
      \rlap{\hbox to \dimen1{\hfil$#1$\hfil}} / \fi}
\ifx\oldchi\undefined \let\oldchi=\chi
  \def\cchi{{\raise 1pt\hbox{$\oldchi$}}} \let\chi=\cchi \fi
  
\def\del{\partial}   

\def\frac#1#2{{\textstyle{#1 \over #2}}}

\def\half{\ifinner {\scriptstyle {1 \over 2}}\else {1 \over 2} \fi}

\def\simge{\rlap{\raise 2pt \hbox{$>$}}{\lower 2pt \hbox{$\sim$}}}
\def\simle{\rlap{\raise 2pt \hbox{$<$}}{\lower 2pt \hbox{$\sim$}}}

\def\vbig#1#2{{\vbigd@men=#2\divide\vbigd@men by 2%
\hbox{$\left#1\vbox to \vbigd@men{}\right.\n@space$}}}

%
%
\newcount\fdefcontre \newcount\fdefcount \newcount\indcount
\newread\filefdef  \newread\fileftmp  \newwrite\filefdef
\newwrite\fileftmp     \def\strip#1*.A {#1}
\def\futuredef#1{\beginfdef
\expandafter\ifx\csname#1\endcsname\relax%
{\immediate\write\fileftmp {#1*.A}
\immediate\write16 {Panda Warning - fdef: macro "#1" on page
\the\pageno \space undefined}
\ifnum\draftnum=0 \expandafter\xdef\csname#1\endcsname{(?)}
\else \expandafter\xdef\csname#1\endcsname{(#1)} \fi
\global\advance\fdefcount by 1}\fi   \csname#1\endcsname}

\def\beginfdef{\ifnum\fdefcontre=0
\immediate\openin\filefdef \jobname.fdef
\immediate\openout\fileftmp \jobname.ftmp
\global\fdefcontre=1  \ifeof\filefdef \immediate\write16 {Panda
WARNING - fdef: file \jobname.fdef not found, run TeX again}
\else \immediate\read\filefdef to\spzzttrra
\global\advance\fdefcount by \spzzttrra
\indcount=0      \loop\ifnum\indcount<\fdefcount
\advance\indcount by 1   \immediate\read\filefdef to\spezttrra
\immediate\read\filefdef to\sppzttrra
\edef\spzzttrra{\expandafter\strip\spezttrra}
\immediate\write\fileftmp {\spzzttrra *.A}
\expandafter\xdef\csname\spzzttrra\endcsname{\sppzttrra}
\repeat \fi \immediate\closein\filefdef \fi}
\def\endfdef{\immediate\closeout\fileftmp   \ifnum\fdefcount>0
\immediate\openin\fileftmp \jobname.ftmp
\immediate\openout\filefdef \jobname.fdef
\immediate\write\filefdef {\the\fdefcount}   \indcount=0
\loop\ifnum\indcount<\fdefcount    \advance\indcount by 1
\immediate\read\fileftmp to\spezttrra
\edef\spzzttrra{\expandafter\strip\spezttrra}
\immediate\write\filefdef{\spzzttrra *.A}
\edef\spezttrra{\string{\csname\spzzttrra\endcsname\string}}
\iwritel\filefdef{\spezttrra}
\repeat  \immediate\closein\fileftmp \immediate\closeout\filefdef
\immediate\write16 {Panda Warning - fdef: Label(s) may have changed,
re-run TeX to get them right}\fi}
\def\iwritel#1#2{\newlinechar=-1
{\newlinechar=`\ \immediate\write#1{#2}}\newlinechar=-1}
\global\fdefcontre=0 \global\fdefcount=0 \global\indcount=0
%
%
\null
%
%
%
%


%
\loadamsmath
%
\pageno=0\baselineskip=14pt
\nopagenumbers{
\line{\hfill CERN-TH.7293/94}
\line{\hfill SWAT/93-94/32}
\line{\hfill\tt hep-th/9407113}
\line{\hfill July 1994}
\ifdoublepage \bjump\bjump\bjump\bjump\else\vfill\fi
\centerline{\capsone Integrable theories that are asymptotically CFT}
\bjump\bjump
\centerline{\scaps Jonathan M.~Evans\footnote{$^{*}$}{Supported by a
fellowship from the EU Human Capital and Mobility programme} and
Timothy J.~Hollowood\footnote{$^{**}$}{On leave from:
Department of Physics, University of Wales, Swansea, SA2
8PP, U.K.}}
\sjump
\sjump
\centerline{\sl CERN-TH, CH-1211 Geneva 23, Switzerland.}
\centerline{\tt evansjm@surya11.cern.ch, hollow@surya11.cern.ch}
\bjump\bjump\bjump
\ifdoublepage
\vfill
\noindent
\line{CERN-TH.7293/94\hfill}
\line{July 1994\hfill}
\eject\null\vfill\fi
\centerline{\capsone ABSTRACT}\sjump
A series of sigma models with torsion are analysed which generate their
mass dynamically but whose ultra-violet fixed points are non-trivial
conformal field theories -- in fact SU(2) WZW models at level $k$.
In contrast to the more familiar situation of asymptotically free theories
in which the fixed points are trivial, the sigma models considered here
may be termed ``asymptotically CFT''.
These theories have previously been conjectured to be quantum integrable;
this is confirmed by postulating a factorizable S-matrix to describe their
infra-red behaviour and then carrying out a stringent test of this proposal.
The test involves coupling the theory to a conserved charge and
evaluating the response of the free-energy both
in perturbation theory to one loop and directly from the S-matrix via
the Thermodynamic Bethe Ansatz with a chemical potential at zero
temperature. Comparison of these results
provides convincing evidence in favour of the proposed S-matrix;
it also yields the universal coefficients of the
beta-function and allows for an evaluation of the mass gap (the ratio of the
physical mass to the $\Lambda$-parameter) to leading order in $1/k$.

\sjump\vfill
\ifdoublepage \else
\noindent
\line{CERN-TH.7293/94\hfill}
\line{July 1994\hfill}\fi
\eject}
\yespagenumbers\pageno=1
%
%
\def\t{\theta}
\def\p{\prime}
\def\d{ {\rm d} }
\def\tr{ {\rm Tr} }
\def\Vol{ V }
\def\zm{ \zeta_M }
\def\Det{ {\rm Det} }
\def\G{ {\cal G} }

\chapter{Introduction}

A theory like QCD with massless quarks in four dimensions has no explicit
mass parameters in its classical Lagrangian; instead a mass scale $\Lambda$ is
generated dynamically at the quantum level.
The quantity $\Lambda$ sets the scale of low-energy physics
so that the masses of all states in the theory, glue-balls, protons,
{\it etc.}, are simply numbers times $\Lambda$.
These numbers are notoriously difficult to extract in QCD, either on
the lattice or analytically.
At energies much greater than $\Lambda$, on the other hand, the theory is
asymptotically free and perturbation theory can be used reliably.
In the language of the renormalization group (RG),
QCD is described by a trajectory
emanating from a fixed point which corresponds to a free theory of gluons and
quarks, the direction of the trajectory being determined by an operator which
is {\it marginally relevant\/}, by which we mean that it is marginal but not
truly marginal.
The fact that the operator is marginal means that
no explicit mass scale is introduced at the fixed point itself,
whilst the fact that it is not truly marginal means that conformal invariance
is broken by the dynamical generation of the scale $\Lambda$ as one moves
away from the fixed point.
The RG trajectory is specified by a running coupling
constant $e(h)$
which depends upon the mass scale scale $h$ being probed and which
in the ultra-violet regime (large $h$) behaves like
$$
{1\over e(h)}=\beta_0\ln(h/\Lambda)+{\beta_1\over\beta_0}
\ln\ln(h/\Lambda)+{\cal O}\left({\ln\ln(h/\Lambda)\over\ln(h/\Lambda)}\right),
\nfr{BET}
which in fact serves to define $\Lambda$ precisely. In the above
$\beta_0$ and $\beta_1$ are universal numbers which appear as the
first two coefficients of the beta-function in perturbation theory.

A more general situation can be envisaged for a theory with dynamical
mass generation, namely the ultra-violet fixed point of the theory,
while necessarily conformally invariant, need not be
free. The purpose of this paper is to analyse such a situation in
two dimensions in which the ultra-violet fixed point is a non-trivial
Conformal Field Theory
(CFT) -- in fact a WZW model. The direction of the RG trajectory is again
determined by some marginally relevant
operator and we say that the theory is ``asymptotically CFT''.

In the case of QCD the main difficulty is the
absence of non-perturbative calculational techniques which can be applied
in the low-energy regime.
In two dimensions, however, there is a rich class of
asymptotically-free theories
which are integrable: the O($N$) sigma models; the
principal chiral models; and the Gross-Neveu models. In these theories
the existence of higher spin conserved charges means that
the S-matrix factorizes, a property which allows in some cases for
its complete determination (see [\Ref{SG}], [\Ref{ORW}]
and [\Ref{KT}] respectively)
yielding an exact description of the low-energy physics.
These integrable theories are therefore particularly
interesting from a theoretical point of view since they provide an
arena in which one can attempt to understand the connection between the
infra-red and ultra-violet regimes.
Such a connection is also important in order to confirm the S-matrices
written down for these models. This is because the S-matrices must,
in the first instance, be regarded as conjectures which
should be tested; in
particular the question of CDD ambiguities
must be resolved.\note{In the case of
the SU$(N)$ chiral
Gross-Neveu model there is a derivation of the S-matrix from first
principles via the Bethe Ansatz [\Ref{AL}].}

In a series of papers ([\Ref{HMN},\Ref{HN}] for the O($N$) sigma
model, [\Ref{BNNW}] for the SU($N$) principal chiral model,
[\Ref{THIII}] for the SO($N$) and Sp($N$) principal chiral models and
[\Ref{FNW}] for the O($N$) Gross-Neveu model) various
authors have used a technique relying on
integrability to relate the infra-red and ultra-violet physics of
families of integrable models, building on the original work of
[\Ref{PW},\Ref{W}].
The idea is to compute a particular physical
quantity -- the free-energy in the presence of a coupling to a conserved
charge -- in two ways: firstly from the S-matrix using
a technique known as the Thermodynamic Bethe Ansatz (TBA) and secondly
from the lagrangian via perturbation theory. For the cases
mentioned above the results of the two calculations are found to be in
perfect agreement in the ultra-violet regime,
thus resolving the problem of CDD ambiguities and,
as a bonus, yielding an exact expression for
the mass gap (the ratio of the physical mass to the
$\Lambda$-parameter). As well as providing a very stringent test of
the form the S-matrix, knowing the mass-gap ratios is interesting in its
own right as they can be compared directly
with the results of lattice simulations.

In this paper we analyse a class of theories which are
asymptotically WZW models
based on the group SU(2) at level $k$ (see {\it e.g.}~[\Ref{EW},\Ref{GO}]).
In accordance with the general situation described above,
the model will correspond to
an RG trajectory defined by the marginally relevant operator
Tr$(J_LJ_R)$ in the WZW theory, where $J_L$ and $J_R$ are the usual
left/right conserved Kac-Moody currents.
This family of examples fits into the general scheme of
``massive current algebras'' set out in [\Ref{DB}].
It is crucial that the theories we consider can also be described
explicitly at the lagrangian level: they are in fact
sigma models with a Wess-Zumino (WZ) term defined on the group
manifold SU(2). This family of lagrangians was first written
down by Balog {\it et al\/} in
[\Ref{BFHP}] who argued further that the resulting theories should
be quantum integrable. What
was not so clear in their work was whether the models would lie in the
class of massive current algebras at the quantum level.
Our strategy for showing that these models do lie in that class, and
in particular that they are quantum integrable,
is to use the exact S-matrices that has been proposed by Ahn {\it et al\/}
[\Ref{LB}] to describe perturbations of WZW models.
We shall then use the
ideas of [\Ref{HMN}--\Ref{FNW}] to carry out a highly non-trivial consistency
check between the lagrangian formulation of [\Ref{BFHP}] and the
S-matrix written down in [\Ref{LB}] in the manner we have already outlined
above. As a by-product we will
extract an expression for the mass gap valid to leading order in $1/k$.

The paper is organized as follows.
In section 2 we discuss the lagrangian for the model,
its current algebra, and its renormalization to one-loop.
In section 3 we write down the S-matrix conjectured to describe the quantum
scattering and in section 4 this is used in conjunction with TBA techniques to
calculate the response of the free-energy to an external field.
Section 5 contains a calculation
of this same quantity in perturbation theory, after which we compare
the expressions to confirm the choice of S-matrix and extract the mass gap
of the model. We conclude with some further remarks in section 6.

\chapter{The lagrangian, current algebra and one-loop renormalization}

The integrable field theories that we shall
investigate are described in two-dimensional Minkowski space-time
(with coordinates $\xi^\mu = (\tau, \sigma)$)
by the lagrangian density [\Ref{BFHP}]
$$\eqalign{
{\cal L}_0={1\over2e^2}&\left\{{1\over x^2-1}\left(\partial_\mu w
\right)^2+
{\beta(w)\over x+1}\left(\partial_\mu
n_a\right)^2+\right.\cr
&\qquad\left.+{1\over
x+1}\left[{1\over\sqrt{x^2-1}}\left({\pi\over2}-w\right)-
\alpha(w)\right]\epsilon_{abc}\epsilon^{\mu\nu}
n_a\partial_\mu n_b\partial_\nu n_c\right\},\cr}
\nfr{LAG}
with
$$
\beta(w)={\cos^2w\over x+\cos2w},\qquad \alpha(w)=\sqrt{x-1\over
x+1}{\sin w\cos w\over x+\cos2w} .
\nfr{AB}
The fields $(w,n_a)$ parameterize the SU(2) group manifold in such a way
that a general group element can be written $g=\cos w+in_a\sigma_a\sin w$ where
the $\sigma_a$'s are the Pauli matrices and
the fields $n_a$ are constrained via $n_1^2+n_2^2+n_3^2=1$.
$e$ and $x$ are coupling constants with $x>1$.

The complicated form of
${\cal L}_0$ requires some explanation.
The most important point is that it ensures that the resulting theory is
classically integrable -- in fact it ensures the existence of
a canonical structure
consisting of two commuting current algebras [\Ref{BFHP}] -- precisely
the structure studied in [\Ref{LB}]. We shall elaborate on this point
below.

The theory has an SU(2) global symmetry generated by transformations
$n_a\mapsto n_a+\epsilon_{abc}q_bn_c$ for parameters $q_b$.
Finite symmetry transformations are given by the adjoint action
$g\mapsto hgh^{-1}$, using some $h\in\,$SU(2).
This is to be contrasted with the principal sigma-model and WZW model
which both have chiral SU(2) $\times$ SU(2) global symmetries.
Our models are invariant under just the diagonal subgroup.

The antisymmetric term in \LAG\ is an example of a Wess-Zumino
(WZ) term and as usual its presence leads to a quantization condition on
coupling constants which is essential in order to obtain a consistent
quantum theory.
In the present case this condition is
$$
{2\pi\over e^2(x+1)\sqrt{x^2-1}}=k\in{\Bbb N}.
\nfr{QK}
One way to derive this is to consider the integral of the curl or exterior
derivative of the WZ term over an arbitrary three-sphere, as in [\Ref{EW}],
and to demand that this always be a multiple of $2 \pi$.
Alternatively one can require that the WZ term itself, although not globally
well-defined, is ambiguous only up to multiples of $2 \pi$.
In our case we can choose the ranges of our coordinates
to be, for example, $0 \leq w < \pi $ with $n_a$ labelling any point on a
two-sphere, which covers SU(2) except for one point.
Then we demand that the integral of
the WZ term should be changed by $2 \pi$ on sending $w \to w + \pi$ which
gives exactly the condition above.\note{The precise relationship
between this criterion and the previous one is quite subtle in the general
case; see {\it e.g.}~[\Ref{A}].}
Yet a third possibility is to appeal to the general representation theory
of Kac-Moody algebras because, as we shall see below, the combination
in \QK\ appears as a central term in the current algebras which
are responsible for the integrability of this model.

All the information concerning the model \LAG\ that
we shall need in the remainder of this paper has now been
set down.
However, in view of the brevity of the presentation in [\Ref{BFHP}]
(and because there appear to be a number of numerical misprints
in the relevant equations which can only be detected after long calculations)
we shall, before proceeding, elaborate on the current algebra structure
which is responsible for the particular form of the Lagrangian \LAG .
We shall also supply some details of the one-loop renormalizability of the
model which were left implicit in [\Ref{BFHP}].

The theory \LAG\ is of the general form
$$
{\cal L}_0={1\over2e^2}\left \{
G_{ij}(\phi)\partial_\mu\phi^i\partial^\mu\phi^j+
\epsilon^{\mu\nu}B_{ij}(\phi)\partial_\mu\phi^i\partial_\nu\phi^j\right \}
\, .
\nfr{GLAG}
where the fields $\phi^i (\xi^\mu)$ describe a map from two-dimensional
Minkowski space-time to some target manifold.
Motivated by the example of WZW models, one can ask
when such a general sigma-model exhibits a classical current algebra.
We restrict attention to the case in which the target manifold is the group
SU(2) and it is convenient for this part of our discussion
to choose antihermitian generators normalized so that
$$
\lambda_a = - {i \over 2} \sigma_a \, , \qquad
[ \lambda_a , \lambda_b ] = \epsilon_{abc} \lambda_c\,,
\efr
which corresponds to choosing the single simple root of SU(2) to
have length one. We shall make no distinction between upper and lower
SU(2) indices.
A natural Ansatz for the light-cone components
$I^a_\pm = I^a_0 \pm I^a_1$ of a current in the SU(2) Lie algebra is
$$
I^a_+ = - {1\over e^2} L^a_i \del_+ \phi^i \, ,
\qquad
I^a_- = - {1\over e^2} R^a_i \del_- \phi^i,
\efr
where $L^a_i$ and $R^a_i$ are vielbeins for the sigma-model metric:
$$
L^a_i L^a_j = R^a_i R^a_j = G_{ij},
\nfr{VB}
The equations of motion following from \GLAG\ ensure that these currents are
conserved $\del_\mu I^{a \mu} = 0$.
It can also be shown by tedious calculation that, with the canonical structure
defined by \GLAG ,
these currents obey a classical (equal-$\tau$) Poisson bracket algebra
$$\eqalign{
\{ I^a_{\pm} (\sigma) , I^b_{\pm} (\sigma^\p) \}
& = \epsilon^{abc} \, ( \, a I^c_{\pm} (\sigma) + b I^c_{\mp} (\sigma) \, )
\delta (\sigma - \sigma^\p) \pm {2\over e^2} \delta^\p (\sigma - \sigma^\p)
\cr
\{ I^a_+ (\sigma) , I^b_- (\sigma^\p) \} & =
- b\, \epsilon^{abc} \, (\, I^c_+ (\sigma) + I^c_- (\sigma) \, )
\delta(\sigma - \sigma^\p),
\cr}
\nfr{calg}
with $a$ and $b$ constants,
provided that the quantities $L^a_i$ and $R^a_i$ satisfy certain conditions.
To express these conditions compactly it is convenient to introduce
differential forms on the group manifold:
$$
L = \lambda_a L^a_i \d \phi^i \, , \qquad
R = \lambda_a R^a_i \d \phi^i.
\efr
Then the current algebra above will hold provided
$$\eqalign{
hL + Rh & = 0 \cr
\d L + a L^2 -  b h^{-1} L^2 h & = 0\cr
\d R + a R^2 - b h R^2 h^{-1} & = 0\cr
3 H = -a \tr L^3 - 3b \tr R L^2 & = a \tr R^3 +  3b \tr L R^2,
\cr}
\nfr{ceqn}
where $h$ is some group-valued function on SU(2) and
$2H = \d B$ is the field-strength three-form corresponding to $B$.
We use the conventions $B = {1 \over 2!} B_{ij} \d \phi^i {\wedge} \d \phi^j$
and $H = {1 \over 3!} H_{ijk} \d \phi^i {\wedge} \d \phi^j {\wedge} \d \phi^k$
for the components of two-forms and three-forms respectively;
as a result
$2H_{ijk} = \del_i B_{jk} + \del_j B_{ki} + \del_k B_{ij}$.

The SU(2) WZW model corresponds to a special solution
of the current algebra conditions above in which
$$
L = g^{-1} \d g
\, , \quad
R = - \d g g^{-1}
\, , \quad
h = g
\, , \quad
a = 1
\, , \quad
b = 0 ,
\efr
and in this case \calg\ clearly collapses to two commuting Kac-Moody algebras.
The action written in \LAG\ corresponds to a slightly more complicated
solution of \ceqn\ which can be motivated as follows.
First consider the Ansatz
$$
L = c (h^{-1} \d h) + \lambda_a n_a f(\rho) \d \rho \, , \quad
R = c ( - \d h h^{-1} ) - \lambda_a n_a f(\rho) \d \rho \, ,
\efr
where $h = \exp (- \lambda_a n_a \rho )$,
$c$ is some constant, $f(\rho)$ is a function to be determined,
and the variable $\rho(w)$ is itself some function of our SU(2) coordinate $w$
which we will fix in a convenient way at the end, after finding a solution.
This seems on the face of it to be a rather redundant
procedure, but it turns out to simplify some technical aspects of the
discussion.
The Ansatz above is clearly a straightforward
modification of the WZW case, and
it is chosen so as to satisfy the first equation in \ceqn\ automatically.
It is not difficult to check that the remaining conditions
in \ceqn\ hold if
$$a = 2x + 1 \, , \quad b = -1 \, , \quad
c = {1 \over 2 (x + 1)} \, , \quad
f(\rho) = {1 \over x+1} {\cos \rho - 1 \over x + \cos \rho}.
\nfr{SOL}
This solution can also be expressed in the form
$$\eqalign{
L & =  \phantom{-} \lambda_a
( \alpha \d n_a  - \beta \epsilon_{abc} n_b \d n_c + \gamma n_a \d \rho )
\cr
R & =  - \lambda_a ( \alpha \d n_a  +  \beta \epsilon_{abc} n_b \d n_c
+ \gamma n_a \d \rho ),
\cr
}
\efr
where the functions $\alpha$, $\beta$ and $\gamma$ are given by
$$
\alpha
= - {\sin \rho \over 2(x+1)} \, , \quad
\beta
= {1 - \cos \rho \over 2(x+1)} \, , \quad
\gamma  = - {1 \over 2(x + \cos \rho)} \, .
\efr
The sigma-model metric and WZ term
are now determined as functions of $\rho$ and $n_a$
by the equations \VB\ and \ceqn\ respectively.

The final step is to relate $\rho$ to $w$, which can be done in such a way
that the expression for the WZ term can be written in closed form.
This is achieved by choosing
$$
{\alpha \over \beta} = - \cot{\rho \over 2} = \sqrt{x-1 \over x+1} \tan w.
\efr
Using this one can deduce the expressions $\alpha (w)$ and $\beta (w)$ given
in \AB\ and, after some effort, one then recovers \LAG .

The current algebra corresponding to the solution \SOL\ above was first
considered by Rajeev [\Ref{R}], who showed that it could be decomposed into two
commuting Kac-Moody algebras. The combinations which achieve this are
$$
J^a_\pm = {1\over 4} \left \{ \,
\left ( {1 \over x+1} + {1 \over \sqrt{x^2 -1} } \right ) I^a_\pm
+ \left ( {1 \over x+1} - {1 \over \sqrt{x^2-1 } } \right ) I^a_\mp
\, \right \},
\efr
obeying
$$\eqalign{
\{ J^a_\pm (\sigma) , J^b_\pm (\sigma^\p) \, \}
& = \epsilon^{abc} J^c_\pm (\sigma) \delta(\sigma - \sigma^\p)
\pm {1 \over 2 e^2 (x+1)\sqrt{x^2 -1} } \delta^\p (\sigma - \sigma^\p) \cr
\{ J^a_\pm (\sigma) , J^b_\mp (\sigma^\p) \, \}
& = 0. \cr
}
\efr
The $\pm$ signs occur in the central terms because these are {\it classical\/}
Kac-Moody algebras, and the
quantization condition \QK\ can now be recovered by comparison
with some standard reference ({\it e.g.}~equation (2.3.14) of [\Ref{GO}]).
Unlike the WZW case, the components of these Kac-Moody currents are
not chirally conserved (although the original current $I^a_\mu$ is
conserved by construction).
Notice, however, that the WZW case can be recovered by taking the limit
$x \to \infty$, $k$ fixed, provided
we rescale the fields appropriately.

Since the theory \LAG\ is a generalized sigma model (a sigma model
with a WZ term) its renormalization group flow can be analysed using the
background field method (see for example [\Ref{RG}]). We can simply
quote the well-known results for the way that the metric and WZ term
in \GLAG\ run with the renormalization scale to one-loop, but we must then
ensure that these equations are indeed consistent with the specific Ansatz
of \LAG . In  our discussion of the current algbera, it was convenient
to keep the coupling constant dependence explicit, but to apply
the general renormalization results of sigma-models it is better
to absorb the coupling constant $e$ into our definitions
of the metric and WZ term by defining
$g_{ij}=G_{ij}/e^2$,
$b_{ij}=B_{ij}/e^2$ and $h_{ijk}=H_{ijk}/e^2$.
The coefficients of the beta-function are calculated in terms of the
generalized curvature ${\hat R}_{ijkl}$ corresponding to the connection
$$
\hat \Gamma_j{}^i{}_k = \Gamma_j{}^i{}_k + h^i{}_{jk}
\efr
which involves the usual Christoffel connection $\Gamma_j{}^i{}_k$
(constructed from the metric $g_{ij}$) modified
by a torsion term.
To one loop one finds [\Ref{RG}] that under the renormalization group
transformation
of the subtraction point $\mu$ the metric and
anti-symmetric field satisfy
$$\mu{\partial g_{ij}\over\partial\mu}=-{1\over2\pi}{\hat R}_{(ij)},\qquad
\mu{\partial b_{ij}\over\partial\mu}=-{1\over2\pi}{\hat R}_{[ij]},
\nfr{FLO}
where ${\hat R}_{ij}={\hat R}^k_{\ ijk}$ is the generalized Ricci tensor.

We now apply these formulae to the theory \LAG. First of all, we
define the coordinates $\t$ and $\psi$ via
$n_a=(\cos\t,\sin\t\cos\psi,\sin\t\sin\psi)$.
In these coordinates the metric has non-zero components
$$
g_{ww}={1\over e^2(x^2-1)},\qquad g_{\t\t}={\beta\over e^2(x+1)},\qquad
g_{\psi\psi}={\beta\over e^2(x+1)}\sin^2\t,
\efr
and the anti-symmetric field has non-zero components
$$
b_{\t\psi}=-b_{\psi\t}={1\over
e^2(x+1)}\left[{1\over\sqrt{x^2-1}}\left({\pi\over2}-w\right)-\alpha\right]
\sin\t.
\efr
{}From these we find that the non-zero components of the generalized
Ricci tensor are
$$\eqalign{
{\hat R}_{ww}&={2\over x^2-1}+4\alpha'\sqrt{x+1\over x-1},\cr
{\hat R}_{\t\t}&=-{2x\over x+1}\beta+2\beta\alpha'\sqrt{x^2-1},\cr
{\hat R}_{\psi\psi}&={\hat R}_{\t\t}\sin^2\t,\cr
{\hat R}_{\t\psi}&=-{\hat
R}_{\psi\t}=2\beta\beta'\sqrt{x^2-1}\sin\t,\cr}
\nfr{CURV}
where the prime denotes a derivative with respect to $w$ at constant $x$.
Using the expression for
the Ricci tensor in \CURV\ in the equations \FLO\
shows that under renormalization group flow the form of the lagrangian
is preserved up to a renormalization of the
coupling constants $e$ and $x$:
$$
\mu{\partial e\over\partial\mu}={1\over2\pi}(1-2x)e^3+{\cal O}(e^5),\qquad
\mu{\partial x\over\partial\mu}={1\over\pi}(x^2-1)e^2+{\cal O}(e^4),
\nfr{RGCC}
and a diffeomorphism of the field $w$ given by
$$
\mu{\partial w\over\partial\mu}=-{1\over\pi}(x^2-1){\cos w\sin w\over
x+\cos2w}e^2+{\cal O}(e^4).
\nfr{DIFF}

These one-loop results \RGCC\ agree with the analysis of
[\Ref{BFHP}]. Notice that to this order $k$ defined in \QK\ is
constant under renormalization group flow as we expect. In
the ultra-violet, $\mu\rightarrow\infty$, $e$ runs to zero and $x$
runs to infinity. In this limit one can easily show that
$$
{\cal L}_0={\cal L}_{\rm WZW}+{k\over8\pi x}{\rm Tr}\left(J_LJ_R\right)+
{\cal O}(1/x^2),
\efr
where ${\cal L}_{\rm WZW}$ is the usual SU(2) WZW lagrangian at level $k$,
and $J_L=g^{-1}\partial_+g$ and $J_R=-(\partial_-g)g^{-1}$ are its
left and right conserved Kac-Moody currents. This expression justifies
our earlier statement that the theories \LAG\ are SU(2) WZW models perturbed
by the operator Tr$(J_LJ_R)$.
It is also easy to see that this perturbation
breaks the chiral SU(2) $\times$ SU(2) symmetry of ${\cal L}_{\rm WZW}$
to the diagonal, or adjoint, SU(2) subgroup mentioned above.

Assuming that $k$ is indeed constant we can eliminate $x$ from \RGCC\
to get the flow equation just involving $e$. Later we shall be
interested in this flow equation for large but finite values of $k$. In this
regime we deduce from \RGCC\ that
$$
\mu{\partial e\over\partial\mu}=-\beta_0e^2-\beta_1e^3-{\cal O}(e^4),
\efr
where
$$
\beta_0=\sqrt{2\over\pi k}\left(1+{\cal O}(1/k)\right),\qquad
\beta_1=-{1\over\pi}\left(1+{\cal O}(1/k)\right).
\nfr{BCO}
Notice that terms coming from higher loops can produce corrections of
lower order in $1/k$ assuming that the coefficient of $e^p$ in
$\mu(\partial e/\partial\mu)$
is polynomial in $x$. It is important to remember that the expressions
\BCO\ are universal.

We have now described in some detail the lagrangian field theory
we wish to study, and in the next section we conjecture an S-matrix
to describe the scattering of states in this theory. We will subsequently
undertake a non-trivial check of the form of the S-matrix
by using the ideas of [\Ref{HMN}-\Ref{FNW}].
To do this we need to
couple the theory to a conserved charge. The idea is to modify the
hamiltonian $H\rightarrow H-hQ$, where $Q$ is a conserved charge
corresponding to some generator of the SU(2) symmetry of the model.
In the Minkowski space lagrangian picture this is
achieved simply by replacing the derivative of $n_a$ in the
time-direction by the ``covariant derivative'':
$$
\partial_0n_a\rightarrow\partial_0n_a+2h\epsilon_{abc}q_bn_c,
\nfr{CCP}
where the $q_a$ are a set of parameters which later we take to be
$q=(1,0,0)$ without loss of generality (due to the SU(2) symmetry).
We will then compute the response of the free-energy per unit volume
$\delta f(h)=f(h)-f(0)$ in the ultra-violet regime in two ways: using
the S-matrix along with thermodynamic Bethe Ansatz techniques and using
conventional perturbation theory.

\chapter{The S-matrix}
Consider, for a moment, the most general way of associating
S-matrices to the Lie algebra
SU($N$). The particles form multiplets associated to the fundamental,
or completely anti-symmetric, representations of the algebra, and each
particle carries in general say $m$
copies of the quantum numbers of that
representation. The general two-body
S-matrix element -- from which all the others may be deduced by
factorization -- has the block form [\Ref{THI},\Ref{THII}]
$$
S^{ab}_{(k_1,k_2,\ldots,k_m)}(\t)=
X^{ab}(\t)S^{ab}_{(k_1)}(\t)\otimes S^{ab}_{(k_2)}(\t)\otimes\cdots
\otimes S^{ab}_{(k_m)}(\t),
\nfr{SGEN}
where factor $S^{ab}_{(k_j)}(\t)$ acts between the $j^{\rm th}$ copies
of the fundamental representations $a$ and $b$ ($a$,
$b=1,2,\ldots,N-1$) and the $k_j$'s are parameters or coupling constants.
The part $X^{ab}(\t)$ is a scalar factor which
ensures that the overall S-matrix has the right analytic structure.
Each block $S^{ab}_{(k)}(\t)$ is invariant under the action of the
quantum loop-group $U_q({\rm SU}(N)\otimes{\Bbb C}[\t,\t^{-1}])$ where
the deformation parameter is $q=-\exp(-i\pi/(N+k))$. In the limit
$k=\infty$ the quantum loop-group reduces to the ordinary loop-group and the
block $S^{ab}_{(\infty)}(\t)$ is invariant under the action of the
group SU($N$) itself. When $k$ is a natural number the blocks are
proportional to RSOS solutions of the Yang-Baxter equation and
$S_{(1)}^{ab}(\t)=1$.

The S-matrix that was proposed in [\Ref{LB}] to describe the
perturbation of the
WZW model of level $k$ is $S^{ab}_{(k,\infty)}(\t)$. For the case of
SU(2) there is only one particle and $X^{11}(\t)=-1$. It is worth
pointing out that this general form subsumes the well-known
S-matrices of the SU($N$) chiral Gross-Neveu model, given by
$S^{ab}_{(\infty)}(\t)\equiv S^{ab}_{(\infty,1)}(\t)$,
and the SU($N$) principal chiral model, given by
$S^{ab}_{(\infty,\infty)}(\t)$. Remarkably, this implies
that the model \LAG\ is
equivalent at the quantum level to the SU$(2)$ chiral
Gross-Neveu and principal
chiral models, for $k=1$ and $k=\infty$, respectively. We shall make a
comment about these equivalences at the end of the paper.

We now write down the S-matrix that is proposed to describe
the theory \LAG. As is conventional, we take the kinematic variable to
be the rapidity difference $\theta$ of the incoming
particles.\note{The velocity and rapidity of a particle are related by
$v={\rm tanh}\t$.} The
S-matrix describes one massive particle with internal quantum
numbers and for the two-body process
it has the product form mentioned above [\Ref{LB}]:
$$
S(\theta)=S_{\rm SU(2)}(\theta)\otimes S^{\rm kink}_{(k)}(\theta),
\nfr{FF}
with $k\in{\Bbb N}$ being identified with \QK.
The product form means that the particle carries
two sets of quantum numbers and
each factor acts on one of the sets only. The first factor $S_{\rm
SU(2)}(\theta)$ is the S-matrix of the SU(2) chiral Gross-Neveu model; hence
the particle transforms in the two-dimensional representation of SU(2)
and the two-body S-matrix elements may be written [\Ref{KT}]
$$
S_{\rm SU(2)}(\theta)={\Gamma\left(1-{\theta\over2\pi i}\right)
\Gamma\left({1\over2}+{\theta\over2\pi i}\right)\over
\Gamma\left(1+{\theta\over2\pi i}\right)
\Gamma\left({1\over2}-{\theta\over2\pi i}\right)}\left[{\Bbb P}_{\rm
t}+
\left({\theta+2\pi i\over\theta-2\pi i}\right){\Bbb
P}_{\rm s}\right],
\efr
where ${\Bbb P}_{\rm t,s}$ indicate the triplet and singlet channels.
This part is equal to $-S^{11}_{(\infty)}(\t)$ as written above
and is invariant under SU(2).

The other factor in \FF\ describes the scattering of kink degrees-of-freedom
carried by the particle. The
particle can either be a kink or an anti-kink which interpolates
between a set $k+1$ vacua $\{1,2,\ldots,k+1\}$ with the following
selection rule: a kink can
connect vacuum $a$ with $a+1$ and an anti-kink $a$ with $a-1$.
The S-matrix is that of soliton scattering in the
restricted sine-Gordon model [\Ref{RSG}] so the S-matrix element
for the process $K_{da}(\theta_1)+K_{ab}(\theta_2)\rightarrow
K_{dc}(\t_2)+K_{cb}(\t_1)$ is
$$\eqalign{
S_{(k)}^{\rm kink}&\pmatrix{a&b\cr d&c\cr}(\t)={u(\t)\over2\pi
i}\left({\sinh(\pi a/p)\sinh(\pi c/p)\over
\sinh(\pi d/p)\sinh(\pi b/p)}\right)^{-\t/2\pi i}\cr
&\times\left\{
\sinh\left({\t\over p}\right)\delta_{db}\left({
\sinh(\pi a/p)\sinh(\pi c/p)\over\sinh(\pi d/p)\sinh(\pi b/p)
}\right)^{1/2}+\sinh\left({i\pi -\t\over p}
\right)\delta_{ac}\right\},\cr}
\efr
where $p=k+2$ and
$$\eqalign{
u(\t)=&\Gamma\left({1\over p}\right)\Gamma\left(1+{i\t\over p
}\right)\Gamma\left(1-{\pi+i\t\over p}\right)\prod_{n=1}^\infty
{R_n(\t)R_n(i\pi-\t)\over R_n(0)R_n(i\pi)}\cr
R_n(\t)=&{\Gamma\left({2n\over p}+{i\t\over\pi p}\right)
\Gamma\left(1+{2n\over p}+{i\t\over\pi p}\right)\over
\Gamma\left({2n+1\over p}+{i\t\over\pi p}\right)
\Gamma\left(1+{2n-1\over p}+{i\t\over\pi p}\right)}.\cr}
\efr

We remark that the form of the S-matrix reflects the form of
the lagrangian: the SU(2) part manifests the SU(2) symmetry of the
model and the kink part describes degrees-of-freedom associated to the
periodic field $w$.

\chapter{The free-energy from the S-matrix}

In this section we evaluate the response of the free-energy
$\delta f(h)$ to the coupling with the charge
directly from the S-matrix. The
technique we use is known as the thermodynamic Bethe Ansatz
[\Ref{TBA}]. In its most general form it leads to an expression for
the free-energy in a cylindrical geometry coupled to a
conserved charge which plays the r\^ole of a chemical potential.
In our case we wish to evaluate the free-energy in the plane,
{\it i.e.\/} at zero temperature.

Consider the (one-dimensional) statistical mechanics of a
gas of particles described by the
S-matrix \FF. Since this theory is integrable, the
number of particles is conserved under interaction
and it is meaningful to consider single
particle energy levels. In a free-theory the energy of these levels
would simply be $\epsilon(\theta)=m\cosh\theta-h$, where $h$ is the
chemical potential, and the free-energy (per unit volume) at zero temperature
would be that of a free one-dimensional relativistic fermion gas:\note{The
fact that particles should be treated as fermions in the thermodynamic Bethe
Ansatz results from that fact that the S-matrix satisfies $S(0)=-1$.}
$$
f(h)={m\over2\pi}\int_{-\t_{\rm F}}^{\t_{\rm F}}d\t\epsilon(\theta)\cosh \t ,
\nfr{FE}
where $\t_{\rm F}$ the Fermi rapidity is determined by the condition
that $\epsilon(\pm\t_{\rm F})=0$.
In our case, the complications are two-fold: the theory is, after all,
not free and furthermore the particles carry internal quantum numbers. As a
result of the
former $\epsilon(\theta)$ now satisfies an integral equation involving
kernels related to the S-matrix of the theory. The
effect of the internal quantum numbers is to couple the energy
$\epsilon(\theta)$ to the ``magnon'' energy levels $\xi_p(\theta)$,
$p=1,2,\ldots,k-1$, and $\zeta_q(\theta)$, $q=1,2,\dots,\infty$; where
the former result from the kink part of the S-matrix and the
latter from the SU(2) part. The free-energy is then still given by
\FE. The equations are known as the
TBA equations and they have been derived at finite
temperature and zero chemical potential for our
S-matrix in [\Ref{ZAM}].\note{The TBA equations for the more general
S-matrices \SGEN\
was considered in [\Ref{THII}].}  At $T=0$ and in the presence of a
chemical potential coupling to the charge of the SU(2) symmetry,
the TBA equations adopt the form
$$\eqalign{
&\epsilon^+(\theta)+R\ast\epsilon^-(\theta)+\sum_{p=1}^{k-1}
a_p^{(k)}\ast\xi_p^+(\theta)
+\sum_{q=1}^\infty a_q^{(\infty)}\ast
\zeta_q^+(\theta)=m\cosh\theta-h,\cr
&\xi_p^-(\theta)+\sum_{q=1}^{k-1}
A^{(k)}_{pq}\ast\xi_q^+(\theta)=a_p^{(k)}\ast\epsilon^-(\theta),\cr
&\zeta_p^-(\theta)+\sum_{q=1}^\infty A^{(\infty)}_{pq}\ast
\zeta_q^+(\theta)=a_p^{(\infty)}\ast\epsilon^-(\theta)-2hp,
}\nfr{TBA}
where star denotes the convolution $f\ast g(\theta)=\int d\theta'
f(\theta-\theta')g(\theta')$ and $f^\pm(\t)$ denote the positive/negative
decomposition of $f(\t)=f^+(\t)+f^-(\t)$, {\it i.e.}
$f^{\pm}(\theta)=f(\theta)$ if $f(\theta)>0$ or $f(\t)<0$,
respectively, being otherwise zero. The kernels in \TBA\ are given by
$$\eqalign{
R(\theta)=&\int_0^{\infty}{dx\over\pi}\cos(\theta x){
\sinh^2(\pi x/2)\over\sinh(\pi kx/2)\sinh(\pi x)}\exp(k\pi
x/2),\cr
A^{(k)}_{pq}(\theta)=&\int_0^\infty{dx\over\pi}\cos(\theta x)
{2\sinh(\pi
px/2)\sinh(\pi(k-q)x/2)\cosh(\pi x/2)\over\sinh(\pi x)
\sinh(\pi x/2)},\cr
a^{(k)}_p(\theta)=&{1\over\pi k}\cdot{\sin(\pi
p/k)\over\cosh(2\theta/k)-\cos(\pi p/k)},
}\efr
for $q\geq p$ ($A_{pq}^{(k)}(\t)=A_{qp}^{(k)}(\t)$).
The dependence of the free-energy (per unit volume) on the chemical
potential is then given by
$$
\delta f(h)=f(h)-f(0)={m\over2\pi}\int_{-\infty}^{\infty}d\theta
\epsilon^-(\theta)\cosh\t.
\efr

The problem before us is to solve the coupled integral equations \TBA.
Our strategy will implicitly assume that the solution of the equations
\TBA\ is unique. Given this the crucial
observation is that $a^{(k)}_p(\theta)$ is a positive kernel; hence
the solution of the TBA equations is
$\xi_p^+(\theta)=\zeta_q^+(\theta)=0$ with
$\xi_p^-(\theta)=a_p^{(k)}\ast\epsilon^-(\theta)$,
$\zeta_p^-(\theta)=a_p^{(\infty)}\ast\epsilon^-(\theta)-2hp$ and
$$
\epsilon^+(\theta)+R\ast\epsilon^-(\theta)=m\cosh\theta-h.
\nfr{STBA}
The solution $\epsilon(\t)$ to \STBA\ is some symmetric convex
function with zeros at the Fermi rapidity $\pm\t_{\rm F}(h)$.
When $h<m$ the system is
below threshold; the external field is too weak to excite any particle states
and hence $\delta f(h)$ is
zero. Beyond the threshold $h=m$ the chemical potential forces
the system into a state where the particles line up their spins with
the external field. From the form of the TBA equations we see that the
external field does not couple to the kink
number and it turns out that the ground-state has total kink number zero.

Notice that the reasoning which led to identifying the solution of the
TBA equations in terms of a particular configuration of the quantum numbers
of the particles was arrived at
from studying the full TBA equations. This is to be contrasted with
the more
heuristic arguments used in [\Ref{HMN}-\Ref{FNW}] leading to the hypothesis
that only one particle-state contributed to the ground-state.

We have arrived at an expression for the free-energy in terms of a
single function $\epsilon(\theta)$ which satisfies the single integral
equation \STBA. This equations are of the same form, but with a
different kernel, as those
of the O($N$) sigma model [\Ref{HMN},\Ref{HN}],
principal chiral models [\Ref{BNNW},\Ref{THIII}] and Gross-Neveu
models [\Ref{FNW}].

It is not possible to solve the equation \STBA\ in closed form;
however, we will be interested in the solution only in the deep ultra-violet
$h\gg m$ for which one can develop a series solution using generalized
Wiener-Hopf techniques [\Ref{HMN},\Ref{JNW}] (see the appendix of
[\Ref{FNW}] for a clear
summary). Rather than explain these techniques we simply follow the
manipulations of [\Ref{FNW}] required to extract the series solution.

The method starts by decomposing the Fourier transform of the kernel
$R(\theta)$ in the following way:
$$
{\sinh^2(\pi x/2)\over\sinh(\pi kx/2)\sinh(\pi
x)}\exp\left({k\pi x/2}\right)={1\over G_+(x)G_-(x)},
\efr
where $G_-(x)=G_+(-x)$ and $G_\pm(x)$ are analytic in the upper/lower
half-planes, respectively. So
$$
G_+(x)=\sqrt{2k}{\Gamma^2\left(1-i{x/2}\right)\over\Gamma\left(
1-i{kx/2}\right)\Gamma\left(1-ix\right)}\exp\left(ixb-i{kx\over2}\ln(-ix)
\right),
\efr
where
$$
b={k\over2}-\ln2-{k\over2}\ln{k\over2}.
\efr
Following [\Ref{FNW}] we now define the function
$\alpha(x)=\exp(2ix\t_{\rm F})G_-(x)/G_+(x)$, where
$\epsilon(\pm\theta_{\rm F})=0$. $\alpha(x)$ has a cut along the
positive imaginary axis and we define $\gamma(\xi)$ in terms of the
discontinuity:
$$
\alpha(i\xi+0)-\alpha(i\xi-0)=-2ie^{-2\xi\t_{\rm F}}\gamma(\xi),
\efr
giving in this case
$$
\gamma(\xi)=\exp\left(-k\xi\ln\xi+2\xi
b\right){\Gamma^2\left(1-{\xi/2}\right)\Gamma\left(1+{k\xi/2}\right)
\Gamma(1+\xi)\over\Gamma^2\left(1+{\xi/2}\right)\Gamma\left(1-{k\xi/
2}\right)\Gamma(1-\xi)}\sin\left({\pi k\xi/2}\right).
\nfr{GEX}
If one consults [\Ref{FNW}] then it soon becomes apparent that
$\gamma(\xi)$ for our model has the same functional form as
a fermion model, rather than a sigma model, namely
$$
\gamma(\xi)=\pi e^{-k\xi\ln\xi}\sum_{n=1}^\infty d_n\xi^n.
\efr
The expansion of the free-energy is given in terms of the quantities
$d_j$. It turns out that $\delta f(h)/h^2$ is power series in the
effective coupling $u=u(h)$ defined through
$$
{1\over u}-k\ln u={1\over z},
\efr
where
$$
{1\over z}=\ln\left[{h^2\over m^2}\left({2G_+(0)\over G_+(i)}\right)^2\right].
\efr
Putting these expression together with the results of [\Ref{FNW}] allows us
to extract the
first few terms in the expansion of the free-energy as a function of $h/m$
$$\eqalign{
&\delta f(h)=\cr
&-{h^2\over2\pi}G_+(0)^2\left\{{1-2d_1 z+2kd_1z^2\ln
z}-2\left[2d_1-\Gamma'(2)kd_1-d_1^2+d_2\right]z^2\right.\cr
&\left.-2k^2d_1z^3\ln^2z+2k\left[4d_1-\Gamma'(3)kd_1-2d_1^2+2d_2\right]z^3\ln
z+{\cal O}(z^3)\right\}.\cr
}\efr
In the above, $\delta f(h)$ is an expansion in terms of the form
$z^m\ln^n z$ where $m>n$. From \GEX\ we find
$$
d_1={k\over2},\quad
d_2=-k\ln2-{k^2\over2}\ln\left({k\over2}\right)+{k^2\over2}\Gamma'(2).
\efr

So putting everything together for the models that we are considering
we find that the free-energy extracted from the S-matrix yields the following
expansion in the ultra-violet:
$$\eqalign{
\delta f(h)=&-{h^2k\over\pi}\left\{1-{k\over2}s+{k^2\over4}s^2\ln s
-{k\over2}\left[1-{1\over2}\ln{64k\over\pi}-{k\over4}-{k\over2}\ln{k\over4}
\right] s^2\right.\cr
&\qquad\left.-{k^3\over8}s^3\ln^2s+{k^2\over2}\left[1-{1\over2}\ln{64k\over
\pi}-{k\over2}-{k\over2}\ln
{k\over4}\right]s^3\ln s+{\cal O}(s^3)\right\},\cr
}\nfr{FES}
where $s^{-1}=\ln(h/m)$.

\chapter{The free-energy from perturbation theory}

In this section, we develop the expansion of the free-energy $\delta f(h)$ in
perturbation theory. We assume, following the discussion to one-loop
in section 2 and [\Ref{BFHP}], that under
renormalization group flow $k$ is constant. Therefore, we may express $1/x$
in terms of $e$:
$$
{1\over x}=\sqrt{k\over2\pi}e+{k\over4\pi}e^2+{\cal O}(e^3),
\nfr{XEXP}
and we shall find that in the ultra-violet $e$ runs to zero and hence
$x$ runs to infinity. The loop expansion parameter is $e^2(x+1)^2$
which is expressed in terms of $e$ as
$$
e^2(x+1)^2={2\pi\over
k}\left(1+\sqrt{k\over2\pi}e+{k\over2\pi}e^2+{\cal O}(e^3)\right).
\nfr{XIE}
This means that
the contributions from higher loops can lead to terms of the same
order in $e$, but their coefficients will be suppressed by
higher powers of $1/k$. Our result to one-loop will therefore be valid in the
large but finite $k$ limit.

The Minkowski space lagrangian in the presence of the chemical potential is
given by substituting \CCP\ in \LAG\ which on Wick rotating to Euclidean space
becomes
$$\eqalign{
&{\cal L}={\cal L}_0
-{2h^2\beta(w)\over e^2(x+1)}\left(1-n_1^2\right)\cr
&+{2h\over e(x+1)}\left\{\beta(w)(n_2\partial_0n_3-n_3\partial_0n_2)
+\left[{1\over\sqrt{x^2-1}}\left({\pi\over2}-w\right)-\alpha(w)\right]
\partial_1n_1\right\}.\cr
}\efr
The ground-state of the system is given by $n_1= w=0$ (modulo some discrete
ambiguity depending on the precise way we parametrize SU(2)).
We wish to calculate the change in the free-energy per-unit-volume
$\delta f(h) $ to one-loop, so it suffices to
expand the lagrangian to quadratic order around the ground
state with $(n_2,n_3)=\sqrt{1-n_1^2}(\cos\psi,\sin\psi)$.
On suitably re-scaling the field $w$ we find
$$\eqalign{
{\cal L}={1\over2e^2(x+1)^2}&\left\{(\partial_\mu w)^2+(\partial_\mu
n_1)^2+(\partial_\mu\psi)^2-{8 hx\over x+1}w\partial_1n_1\right.\cr
&\left.-4h^2+4h^2n_1^2+4h^2w^2\left({x-1\over
x+1}\right)^2+\cdots\right\},\cr
}\efr
where the ellipsis represent the interaction.

We can simply read off the
tree-level contribution to $\delta f(h)$:
$$
\delta f(h)_0=-{2h^2\over
e^2(x+1)^2}=-{h^2k\over\pi}\left\{1-\sqrt{k\over2\pi}e+{\cal O}(e^3)\right\}.
\efr
The one-loop contribution is, from the quadratic Euclidean lagrangian
written above,
$$
\delta f(h)_1 = {1 \over 2} \ln \Det \left \{ {M \over e^2 (x+1)^2} \right \},
\nfr{ONEL}
where the operator $M$ acts on $(n_1 , w)$ according to
$$
M ={1 \over \mu^2} \pmatrix{ -\del^2+4h^2 & { 4 h x\over x+1} \del_1 \cr
-{4hx \over x+1} \del_1 & -\del^2+4h^2
\left({x-1\over x+1}\right)^2\cr} \, .
\efr
Here $\mu$ is a mass-scale introduced to make the eigenvalues
of $M$ dimensionless.
Note also that the field $\psi$ is completely decoupled to this order
in the loop expansion.

The operator above is rather unconventional in nature and it proves convenient
to evaluate its determinant using zeta-function techniques.
Some basic facts concerning these methods, together with their application to
operators of the above type, are summarized in the appendix.
The important point for our purposes is that
$$
\ln \Det\, M = - \zm' (0),
\nfr{ZEQN}
where the zeta-function $\zm (s)$ corresponding to the operator $M$ can be
represented by
$$
\zm (s) = { 2 V \mu^{2s} \over \Gamma(s)} \int_0^\infty dt \, t^{s-1}
\int {d^2 p \over (2 \pi)^2}
e^{ -t (p^2 + \lambda)} \cosh\left( t ( \eta^2 p_1^2 + \rho^2
)^{1/2}\right)\, ,
\nfr{INT}
with
$$
\lambda = { 4h^2(x^2 + 1) \over (x+1)^2} \, , \qquad
\rho = { 8 h^2 x \over (x + 1)^2} \, , \qquad
\eta = {4hx \over x+1}.
\nfr{CONSTS}
The factor $V$ denotes formally the volume of two-dimensional
spacetime (strictly speaking
this should be dealt with using some explicit infra-red
regularization but these details are irrelevant for our purposes).
Unfortunately the integral above cannot be evaluated in closed form,
but we shall sketch below how it can be successfully expanded in powers of
$1/x$ to the order which we need.
(Recall that the ultra-violet limit of our models is $x \to \infty$ with $k$
fixed.)

The strategy is to expand the cosh factor in \INT\
as a series and collect terms of a given power in $p_1$ so that the momentum
integrals can be evaluated.
The $t$ integrals can then be expressed in terms of $\Gamma$-functions,
and one obtains the result
$$
\zm (s) = {V \mu^{2s} \over 2 \pi \Gamma (s)}
\sum_{m = -1}^\infty C_m \lambda^{-(s+m)} \Gamma (s + m) \, ,
\efr
where the coefficients are given by the rather complicated expressions
$$
C_m = \sum_{ {m=2n-r-1 \atop 0\leq r\leq n\leq \infty } }
{ (2r)! n! \over (2n)! (r!)^2 (n-r)! 2^{2r}} \eta^{2r} \rho^{2(n-r)} \, .
\nfr{UGH}
It is not obvious how to sum the series above completely, but we note
that $\rho$ in these expressions is
${\cal O}(1/x)$ which will enable us to simplify things presently.

On differentiating and setting $s=0$ we find
$$
\zm'(0) = {V\over 2 \pi} \left \{
- \lambda  + \left (\lambda - {\eta^2 \over 4} \right )
\ln {\lambda \over \mu^2}
+ \sum_{m=1}^\infty C_m \lambda^{-m} \Gamma (m)
\right \}.
\nfr{ANS}
At this stage our result is still exact, but we must now consider its
behaviour as a power series in $\rho$ (and hence $1/x$) to make
further progress. The leading contribution is ${\cal O} (\rho^0)$;
extracting all such terms from the expressions \UGH\ for $C_m$ with $m \geq 1$
gives us an infinite series which can be summed\note{
$\sum_{n=2}^{\infty} {1 \over n (n-1)} y^n = (1 -y) ( \ln(1-y) - 1) + 1$
for suitable $y$.}
to yield
$$
{\Vol \over 2 \pi} \lambda \left \{
\left( 1  - {\eta^2 \over 4 \lambda }\right)
\left( \ln \left( 1 - {\eta^2 \over 4 \lambda} \right) - 1 \right )
+ 1 \right \}.
\nfr{SUMA}
The next contributions are ${\cal O} (\rho^2)$ and again the series resulting
from the coefficients $C_m$ with $m\geq1$ can be summed\note{
$\sum_{n=1}^\infty {1 \over 2n -1} z^{2n -1} = {1 \over 2} ( \ln (1 + z ) -
\ln (1 - z ) )$ for suitable $z$.}
to give
$$
{V \over 2 \pi} {\rho^2 \over 2 \eta \sqrt{\lambda} }
\left \{ \ln \left( 1 + {\eta \over 2 \sqrt \lambda} \right )
- \ln \left( 1 - {\eta \over 2 \sqrt \lambda} \right ) \right \}.
\nfr{SUMB}
The other contributions to \ANS\ are ${\cal O}(\rho^4) = {\cal O}(1/x^4)$
and we neglect them.
Now we substitute \SUMA\ and \SUMB\ in \ANS\ and expand each of
$\lambda$, $\rho$ and $\eta$ in powers of $1/x$ using \CONSTS .
It turns out that although \SUMA\ is ${\cal O} (\rho^0)$ and
$\rho = {\cal O} (1/x)$ the particular
form of the expressions for $\lambda$ and $\eta$ imply that \SUMA\ is,
more precisely, ${\cal O} (1/x^2)$. The final result is
$$
\zm' (0) = {\Vol \over 2 \pi} {4 h^2 \over x^2}
\left [ \ln {16 h^2 \over \mu^2} - 1 \right] + {\cal O} (1/x^3).
\efr

Now to obtain the required determinant in \ONEL\ all we need do is restore
the factor $e^2 (x+1)^2$
(which we dropped in \ZEQN\
for simplicity) by rescaling $\mu$
and use the consequence \XEXP\ of the
quantization condition \QK\ to eliminate
$x$ in favour of $e$ and $k$.
This gives a final result
for the change in the one-loop contribution to the free-energy per unit
volume of
$$
\delta f(h)_1={h^2k\over\pi}\left\{{e^2\over2\pi}
\left(1-\ln{8kh^2\over\pi\mu^2}\right)+{\cal O}(e^3)\right\}.
\efr

The change in the free-energy $\delta f_0+\delta f_1$
which we have just calculated
is a renormalization group invariant quantity,
{\it i.e.} it is independent of $\mu$ when the
coupling constant runs with $\mu$.
This fact allows us to determine the running of the coupling, as expressed
by the beta-function
$$
\mu{\partial e\over\partial\mu}=-\sqrt{2\over\pi
k}e^2-\beta_1e^3+{\cal O}(e^4),
\nfr{BF}
although the coefficient $\beta_1$ is not determined to the order that
we are working. Since the first two coefficients of the beta-function
are universal, it is comforting that \BF\ is consistent with the
calculation of the beta-function via the background field method
\BCO. It is convenient to use the fact that
the free-energy is independent of $\mu$ to set $\mu=h$
and then, integrating \BF , one obtains
$$
e(h)=\sqrt{\pi k\over2}\xi+\beta_1\left({\pi k\over2}\right)^{3/2}\xi^2\ln\xi
+\beta_1^2\left({\pi
k\over2}\right)^{5/2}\xi^3\left(\ln^2\xi+\ln\xi\right)
+{\cal O}(\xi^3),
\efr
where $\xi^{-1}=\ln(h/\Lambda_\zeta)$ is defined in terms of
the $\Lambda$-parameter of the zeta-function regularization scheme
and the expansion is in terms of the form $\xi^m\ln^n\xi$ with $m>n$.

Hence to one-loop we deduce
$$\eqalign{
\delta f(h)=&-{h^2k\over\pi}\left\{1-{k\over2}\xi-{\pi
k^2\beta_1\over4}\xi^2\ln\xi-{k\over4}\left(1-\ln{8k\over\pi}\right)\xi^2
\right.\cr &\quad\left.-{\pi^2k^3\beta_1^2\over8}\xi^3\ln^2\xi-{\pi
k^2\beta_1\over4}
\left({\pi k\beta_1\over 2}+1-\ln{8k\over\pi}\right)\xi^3\ln\xi
+{\cal O}(\xi^3)\right\}
}\nfr{FEP}
The effect of higher loops would be to introduce corrections at the same
order in $\xi$ but with coefficients which are suppressed by powers of $1/k$.

We now reach our main result: at the order to which we are working
the two expansions \FES\ and \FEP\ are consistent, a fact which provides a
highly non-trivial check of the conjectured S-matrix.
Furthermore, by comparing the
two expressions we can extract the mass-gap ratio for large $k$:
$$
\ln {m\over\Lambda_\zeta}=
-{1\over2}+{3\over2}\ln2+{k\over4}+{k\over2}\ln{k\over4} +
{\cal O}(1/k).
\nfr{MG}
We also deduce that the second coefficient of the beta-function is, for
large $k$, simply $\beta_1=-1/\pi+{\cal O}(1/k)$, a result which is in
perfect agreement with the second coefficient of the beta-function
computed directly using the background field method \BCO.

\chapter{Conclusions}

We have investigated a series of theories that generate their mass
dynamically but which are asymptotically (in the ultra-violet limit)
non-trivial CFTs. The
theories are in addition integrable, a property
implying that their S-matrices factorize, allowing us to
conjecture a form for these S-matrices.
The calculations of the free energy which we carried out provide
a highly non-trivial check on the form of the S-matrices since,
as pointed out in [\Ref{BNNW}], the addition of any CDD factors
would drastically alter the thermodynamics of the system and
consequently destroy the remarkable consistency between the TBA calculation
and the perturbative result. We can conclude with some confidence therefore
that the proposed S-matrices are correct and that the classical integrability
of these models extends to the quantum regime. Notice that the
S-matrix has a quantum group symmetry, an invariance which does not
seem to be manifested at the lagrangian level in any simple fashion.

It is worth pointing out that the leading order behaviour
$-h^2k/\pi$
of the free-energy near the ultra-violet fixed-point
can easily be deduced from the knowledge that
in this limit the theory is an SU(2) WZW model of level $k$. At the fixed
point the chemical potential couples to
the combination ${\tilde J}={\tilde J}_L+{\tilde J}_R$, where ${\tilde
J}_L$ and ${\tilde J}_R$ are the left and
right currents of a U(1) subalgebra of the SU(2) current algebra.
Following [\Ref{FI}], the response of the free-energy in a finite
volume $V$ is given by
$$
\delta f(h)=-{h^2\over2V}\int{d^2z\over2\pi}\int{d^2w\over2\pi}\langle\tilde
J(z)\tilde J(w)\rangle,
\efr
where the expectation value is evaluated in the WZW model.
The final result is proportional to the anomaly in the U(1) currents, however
one must take careful account of the form of the operator products in
a finite volume [\Ref{FI}]:
$$\eqalign{
\langle\tilde J_L(z)\tilde J_L(w)\rangle={k\over(z-w)^2}+{2\pi
k\over V},&\qquad
\langle\tilde J_R(z)\tilde J_R(w)\rangle={k\over(\overline z-\overline
w)^2}+{2\pi k\over V},\cr
\langle\tilde J_L(z)\tilde J_R(w)\rangle&=2\pi k\delta^{(2)}
(z-w)-{2\pi k\over V}.\cr}
\efr
It is then straightforward to to extract [\Ref{FI}]
$$
\delta f(h)=-{h^2 k\over\pi},
\efr
in agreement with the leading order term in \FES\ and \FEP.
Obviously one could
extend this calculation away from the fixed-point by perturbation
theory and hope to reproduce the series \FEP; a calculation which
would be interesting since it would be valid not just in the large $k$
regime.

We should also emphasize a remarkable consequence of the equivalence
of the lagrangian and S-matrix descriptions we have established in this paper:
namely that the
field theory \LAG\ for $k=1$ and $k=\infty$ is quantum equivalent to
the SU$(2)$ chiral Gross-Neveu model and principal chiral model,
respectively (since our Ansatz for the S-matrix then reduces to these
well-known cases).
Let us consider the latter equivalence in more detail. In
taking the limit $k\rightarrow\infty$, at fixed $e$,
it is necessary to introduce the field
$r=((\pi/2)-w)/\sqrt{x^2-1}$. The lagrangian then has a well-defined limit
$$
{\cal L}_0={1\over2 e^2}\left\{\left(\partial_\mu
r\right)^2+{r^2\over1+4r^2}\left(\partial_\mu n_a\right)^2+
{2r^3\over1+4r^2}\epsilon_{abc}\epsilon^{\mu\nu}n_a\partial_\mu n_b
\partial_\nu n_c\right\}.
\efr
With a simple change of variables $\phi_a=r n_a$, this may be written
$$
{\cal L}_0={1\over2e^2}\left({1\over1+4\phi^2}\right)
\left\{\left(\delta_{ab}+4\phi_a\phi_b\right)\partial_\mu\phi_a
\partial^\mu\phi_b+2\epsilon_{abc}\epsilon^{\mu\nu}\phi_a\partial_\mu
\phi_b\partial_\nu\phi_c\right\}.
\efr
This lagrangian is actually the non-abelian dual of the SU$(2)$
principal chiral model [\Ref{NAD}] and hence it is indeed known to be quantum
equivalent to it. It would be interesting to show, in a similar spirit,
that \LAG\ with $k=1$ was a bosonized form of the SU$(2)$
chiral Gross-Neveu model.

Finally, it would clearly be interesting to extend the various results
above to larger groups and
also to S-matrices of the more general form \SGEN.

TJH would like to thank Michel Bauer for many interesting discussions.

\appendix{: Some basic facts about zeta-functions}

The zeta-function $\zm$ associated with an operator $M$ with
eigenvalues $\lambda_i$ can be thought of formally as
$$
\zm (s) = \sum_i \lambda^{-s}_i.
\nfr{DEFI}
It is not immediately obvious when this formula
makes sense beyond the simplest circumstances in which
$M$ is finite-dimensional with $\lambda_i > 0$ and Re$(s)$
sufficiently large.
By comparison with this simplest case, however, it should seem reasonable
that when $\zm$ exists it can be used to calculate
the dimensions and determinant of $M$ via
$$
\dim M = \zm (0) \, , \qquad \Det\, M = - \zm' (0).
\nfr{DDEF}
To make use of this we need some means of calculating $\zm$ (without
first finding all its eigenvalues and using \DEFI !) and this is
provided by the heat-kernel representation.
The key is the observation that
\DEFI\ is formally equivalent to
$$
\zm (s) = {1 \over \Gamma (s)} \int_0^\infty dt \, t^{s-1} \tr\, \G (t),
\nfr{ZDEF}
where $\G (t) = \exp (- t M)$.
Then we note that $\G (t) = \exp(-t M)$ can also be characterized by the
`heat' equation and boundary condition
$$
{\del \G \over \del t} = - M \G \, , \qquad \G(0) = 1,
\nfr{HKER}
(where in the boundary condition 1 is the identity operator on the
relevant space). Now we can dispense with \DEFI\ entirely
by adopting \ZDEF\ and \HKER\ as a definition of $\zm (s)$ for any given
(suitably well-behaved) operator $M$ and we can take \DDEF\ as
providing definitions of the dimension and determinant of $M$.

The case of interest for us is an operator in two Euclidean space dimensions
of the form
$$
M= {1 \over \mu^2} \pmatrix{ -\del^2 + a^2 & c \del_1 \cr - c \del_1 &
-\del^2 + b^2 }.
\efr
A solution of \HKER\ can be found by Fourier transforming to momentum
space:
$$
\G (\xi , t) = \int {d^2 p \over (2 \pi)^2} e^{i p\cdot\xi} e^{- t M(p)}
\, , \qquad M(p) =
{1 \over \mu^2} \pmatrix{ p^2 + a^2 & icp_1 \cr - ic p_1 &
p^2 + b^2 \cr } .
\nfr{GSOL}
After substituting in \ZDEF\ we can evaluate the functional part of the
trace immediately to obtain a factor $V$ which is the two-dimensional
space-time-volume. This leaves us still with a matrix trace
$$
\tr\, e^{-t M(p)} = 2 e^{- (t/ \mu^2) (p^2 + \lambda)} \cosh\left((t/ \mu^2)
(c^2 p_1^2 + \rho^2 )^{1/2}\right),
\efr
where
$$
\lambda = {1 \over 2} (a^2 + b^2) \, , \qquad
\rho = {1 \over 2} (a^2 - b^2) \, .
\efr
On combining this with \ZDEF\ and \GSOL\ and re-scaling $t$ we obtain
the expression \INT\ in the text with $\eta =c$.

\references

\beginref
\Rref{ORW}{E. Ogievetsky, N. Reshetikhin and P. Wiegmann, Nucl. Phys.
{\bf B280} (1987) 45}
\Rref{FNW}{P. Forg\'acs, F. Niedermayer and P. Weisz, Nucl. Phys. {\bf
B367} (1991) 123}
\Rref{HN}{P. Hasenfratz and F. Niedermayer, Phys. Lett. {\bf B245}
(1990) 529}
\Rref{BNNW}{J. Balog, S. Naik, F. Niedermayer and P. Weisz,
Phys. Rev. Lett. {\bf 69} (1992) 873;
\newline
S. Naik, Nucl. Phys. {\bf B} (Proc. Suppl.) {\bf 30} (1993) 232}
\Rref{HMN}{P. Hasenfratz, M. Maggiore and F. Niedermayer, Phys. Lett.
{\bf B245} (1990) 522}
\Rref{JNW}{G. Japaridze, A. Nersesyan and P. Wiegmann, Nucl. Phys.
{\bf B230} (1984) 511}
\Rref{SG}{A.B. Zamolodchikov and Al. B. Zamolodchikov, Ann. Phys.
{\bf120} (1979) 253}
\Rref{BFHP}{J. Balog, P.Forg\'acs, Z. Horv\'ath and L. Palla, Phys.
Lett. {\bf B324} (1994) 403}
\Rref{THI}{T.J. Hollowood, Nucl. Phys. {\bf B414} (1994) 379}
\Rref{THII}{T.J. Hollowood, Phys. Lett. {\bf B230} (1994) 43}
\Rref{THIII}{T.J. Hollowood, Phys. Lett. {\bf B329} (1994) 450}
\Rref{RSG}{A. LeClair, Phys. Lett. {\bf B227} (1989) 417\newline
D. Bernard and A. LeClair, Nucl. Phys. {\bf B340} (1990) 721}
\Rref{TBA}{E.H. Lieb and W. Liniger, Phys. Rev. {\bf130} (1963) 1605\newline
Al.B. Zamolodchikov, Nucl. Phys. {\bf B342} (1990) 695}
\Rref{FI}{P. Fendley and K. Intriligator, Phys. Lett. {\bf B319}
(1993) 132}
\Rref{LB}{C. Ahn, D. Bernard and A. LeClair, Nucl. Phys. {\bf B346}
(1990) 409}
\Rref{ZAM}{Al.B. Zamolodchikov, Nucl. Phys. {\bf B366} (1991) 122}
\Rref{PW}{A. Polyakov and P.B. Wiegmann, Phys. Lett. {\bf B131} (1983)
121}
\Rref{W}{P.B. Wiegmann, Phys. Lett. {\bf B141} (1984) 217}
\Rref{KT}{B. Berg and P. Weisz, Nucl. Phys. {\bf B146} (1979)
205\newline
V. Kurak and J.A. Swieca, Phys. Lett. {\bf B82} (1979) 289\newline
M. Karowski and H.J. Thun, Nucl. Phys. {\bf B190} (1981)
61}
\Rref{AL}{N. Andrei and J.H. Lowenstein, Phys. Lett. {\bf B90} (1980) 106}
\Rref{RG}{D. Friedan, Ann. Phys. {\bf 163} (1985) 318\newline
T.L. Curtright and C.K. Zachos, Phys. Rev. Lett. {\bf 53} (1984)
1799\newline
E. Braaten, T.L. Curtright and C.K. Zachos, Nucl. Phys. {\bf B260}
(1985) 630\newline
S. Mukhi, Nucl. Phys. {\bf B264} (1986) 640\newline
C.M. Hull and P.K. Townsend, Phys. Lett. {\bf B191} (1987) 115\newline
D. Zanon, Phys. Lett. {\bf B191} (1987) 363\newline
D.R.T. Jones, Phys. Lett. {\bf B192} (1987) 391\newline
H. Osborn, Ann. of Phys. {\bf200} (1990) 1}
\Rref{DB}{D. Bernard, Commun. Math. Phys. {\bf137} (1991) 191}
\Rref{R}{S.~Rajeev, Phys.~Lett.~{\bf B217} (1989) 123}
\Rref{A}{O.~Alvarez, Commun.~Math.~Phys.~{\bf 100} (1985) 279;
R.~Rohm and E.~Witten, Ann.~Phys.~{\bf 170} (1986) 454}
\Rref{GO}{P.~Goddard and D.~Olive, Int.~J.~Mod.~Phys.~{\bf A1} (1986) 303}
\Rref{EW}{E.~Witten, Commun.~Math.~Phys.~{\bf 92} (1984) 455}
\Rref{NAD}{B. Fridling and A. Jevicki, Phys. Lett. {\bf B134} (1984) 70\newline
E. Fradkin and A. Tseytlin, Ann. Phys. {\bf 162} (1985) 31\newline
C.K.Zachos and T.L. Curtright, Phys. Rev. {\bf D49} (1994) 5408
}
\endref
\ciao